\documentclass[twocolumn]{aastex631}

\usepackage{xspace}
\usepackage{graphicx}
\usepackage{lineno}
\usepackage{ulem}

\newcommand{\tess}{\textit{TESS}\xspace}

\newcommand{\kepler}{\textit{Kepler}\xspace}

\newcommand{\harps}{\textit{HARPS}\xspace}
\newcommand{\serval}{\textsc{serval}\xspace}

\newcommand{\pyaneti}{\texttt{pyaneti}\xspace}

\shorttitle{Company for the ultra-high density, ultra-short period sub-Earth GJ\,367\,b}
\shortauthors{Goffo et al.}

\graphicspath{{./}{figures/}}

\begin{document}

\title{Company for the ultra-high density, ultra-short period sub-Earth GJ\,367\,b:\\
discovery of two additional low-mass planets at 11.5 and 34 days\protect\footnote{Based on observations made with the ESO-3.6 m telescope at La Silla Observatory under programs 1102.C-0923 and 106.21TJ.001.}}

\author[0000-0001-9670-961X]{Elisa Goffo}
\thanks{elisa.goffo@unito.it \\
elisa@tls-tautenburg.de}
\affiliation{Dipartimento di Fisica, Universit\'a degli Studi di Torino, via Pietro Giuria 1, I-10125, Torino, Italy}
\affiliation{Th\"uringer Landessternwarte Tautenburg, Sternwarte 5, D-07778 Tautenburg, Germany}
\author[0000-0001-8627-9628]{Davide Gandolfi}
\affiliation{Dipartimento di Fisica, Universit\'a degli Studi di Torino, via Pietro Giuria 1, I-10125, Torino, Italy}
\author[0000-0003-1628-4231]{Jo Ann Egger}
\affiliation{Physikalisches Institut, University of Bern, Gesellschaftsstrasse 6, 3012 Bern, Switzerland}
\author[0000-0002-2086-3642]{Alexander J. Mustill}
\affiliation{Lund Observatory, Division of Astrophysics, Department of Physics, Lund University, Box 43, SE-221 00 Lund, Sweden}
\affiliation{Lund Observatory, Department of Astronomy and Theoretical Physics, Lund University, Box 43, SE-221 00 Lund, Sweden}
\author[0000-0003-1762-8235]{Simon H. Albrecht}
\affiliation{Stellar Astrophysics Centre, Department of Physics and Astronomy, Aarhus University, Ny Munkegade 120, DK-8000 Aarhus C, Denmark}
\author[0000-0003-3618-7535]{Teruyuki Hirano}
\affiliation{Astrobiology Center, 2-21-1 Osawa, Mitaka, Tokyo 181-8588, Japan}
\affiliation{National Astronomical Observatory of Japan, 2-21-1 Osawa, Mitaka, Tokyo 181-8588, Japan}
\author[0000-0003-3061-4591]{Oleg Kochukhov}
\affiliation{Department of Physics and Astronomy, Uppsala University, Box 516, SE-75120 Uppsala, Sweden}
\author[0000-0002-8462-515X]{Nicola Astudillo-Defru}
\affiliation{Departamento de Matem\'atica y F\'isica Aplicadas, Universidad Cat\'olica de la Sant\'isima Concepci\'on, Alonso de Rivera 2850, Concepci\'on, Chile}
\author[0000-0003-0563-0493]{Oscar Barragan}
\affiliation{Sub-department of Astrophysics, Department of Physics, University of Oxford, Oxford, OX1 3RH, UK}
\author[0000-0001-9211-3691]{Luisa M. Serrano}
\affiliation{Dipartimento di Fisica, Universit\'a degli Studi di Torino, via Pietro Giuria 1, I-10125, Torino, Italy}
\author[0000-0002-3404-8358]{Artie P. Hatzes}
\affiliation{Th\"uringer Landessternwarte Tautenburg, Sternwarte 5, D-07778 Tautenburg, Germany}
\author[0000-0002-4644-8818]{Yann Alibert}
\affiliation{Physikalisches Institut, University of Bern, Gesellschaftsstrasse 6, 3012 Bern, Switzerland}
\affiliation{Center for Space and Habitability, University of Bern, Gesellsschaftsstr. 6 CH3012, Bern, Switzerland}
\author[0000-0002-9130-6747]{Eike Guenther}
\affiliation{Th\"uringer Landessternwarte Tautenburg, Sternwarte 5, D-07778 Tautenburg, Germany}
\author[0000-0002-8958-0683]{Fei Dai}
\thanks{NASA Sagan Fellow.}
\affiliation{Division of Geological and Planetary Sciences,
1200 E California Blvd, Pasadena, CA, 91125, USA}
\affiliation{Department of Astronomy, California Institute of Technology, Pasadena, CA 91125, USA}
\author[0000-0002-9910-6088]{Kristine W. F. Lam}
\affiliation{Institute of Planetary Research, German Aerospace Center (DLR), Rutherfordstrasse 2, D-12489 Berlin, Germany}
\author[0000-0001-6803-9698]{Szil\'ard Csizmadia}
\affiliation{Institute of Planetary Research, German Aerospace Center (DLR), Rutherfordstrasse 2, D-12489 Berlin, Germany}
\author[0000-0002-2386-4341]{Alexis M. S. Smith}
\affiliation{Institute of Planetary Research, German Aerospace Center (DLR), Rutherfordstrasse 2, D-12489 Berlin, Germany}
\author[0000-0003-4426-9530]{Luca Fossati}
\affiliation{Space Research Institute, Austrian Academy of Sciences, Schmiedlstrasse 6, A-8042, Graz, Austria}
\author[0000-0002-4671-2957]{Rafael Luque}
\affiliation{Department of Astronomy and Astrophysics, University of Chicago, Chicago, IL 60637, USA}
\author[0000-0003-0650-5723]{Florian Rodler}
\affiliation{European Southern Observatory, Alonso de Cordova 3107, Vitacura, Santiago de Chile, Chile}
\author[0000-0003-1687-3271]{Mark L. Winther}
\affiliation{Stellar Astrophysics Centre, Department of Physics and Astronomy, Aarhus University, Ny Munkegade 120, DK-8000 Aarhus C, Denmark}
\author[0000-0001-9234-430X]{Jakob L. R{\o}rsted}
\affiliation{Stellar Astrophysics Centre, Department of Physics and Astronomy, Aarhus University, Ny Munkegade 120, DK-8000 Aarhus C, Denmark}
\author{Javier Alarcon}
\affiliation{European Southern Observatory, Alonso de Cordova 3107, Vitacura, Santiago de Chile, Chile}
\author{Xavier Bonfils}
\affiliation{Universit\'e Grenoble Alpes, CNRS, IPAG, 38000 Grenoble, France}
\author[0000-0001-9662-3496]{William D. Cochran}
\affiliation{McDonald Observatory and Center for Planetary Systems Habitability, The University of Texas, Austin Texas USA}
\author[0000-0003-0047-4241]{Hans J. Deeg}
\affiliation{Instituto de Astrof\'isica de Canarias (IAC), E-38205 La Laguna, Tenerife, Spain}
\affiliation{Departamento de Astrof\'isica, Universidad de La Laguna (ULL), E-38206 La Laguna, Tenerife, Spain}
\author[0000-0002-4715-9460]{Jon M. Jenkins}
\affiliation{NASA Ames Research Center, Moffett Field, CA 94035, USA}
\author[0000-0002-0076-6239]{Judith Korth}
\affiliation{Lund Observatory, Division of Astrophysics, Department of Physics, Lund University, Box 43, SE-221 00 Lund, Sweden}
\author[0000-0002-4881-3620]{John H. Livingston}
\affiliation{Astrobiology Center, 2-21-1 Osawa, Mitaka, Tokyo 181-8588, Japan}
\affiliation{National Astronomical Observatory of Japan, 2-21-1 Osawa, Mitaka, Tokyo 181-8588, Japan}
\affiliation{Department of Astronomical Science, School of Physical Sciences, The Graduate University for Advanced Studies (SOKENDAI), 2-21-1, Osawa, Mitaka, Tokyo, 181-8588, Japan}
\author[0000-0002-7500-7173]{Annabella Meech}
\affiliation{Department of Physics, University of Oxford, Keble Road, Oxford, OX1 3RH, UK}
\author[0000-0001-9087-1245]{Felipe Murgas}
\affiliation{Instituto de Astrof\'isica de Canarias (IAC), E-38205 La Laguna, Tenerife, Spain}
\affiliation{Departamento de Astrof\'isica, Universidad de La Laguna (ULL), E-38206 La Laguna, Tenerife, Spain}
\author[0000-0003-2066-8959]{Jaume Orell-Miquel}
\affiliation{Instituto de Astrof\'isica de Canarias (IAC), E-38205 La Laguna, Tenerife, Spain}
\affiliation{Departamento de Astrof\'isica, Universidad de La Laguna (ULL), E-38206 La Laguna, Tenerife, Spain}
\author[0000-0002-4143-4767]{Hannah L. M. Osborne}
\affiliation{Mullard Space Science Laboratory, University College London, Holmbury St Mary, Dorking, Surrey RH5 6NT, UK}
\author[0000-0003-0987-1593]{Enric Palle}
\affiliation{Instituto de Astrof\'isica de Canarias (IAC), E-38205 La Laguna, Tenerife, Spain}
\affiliation{Departamento de Astrof\'isica, Universidad de La Laguna (ULL), E-38206 La Laguna, Tenerife, Spain}
\author[0000-0003-1257-5146]{Carina M. Persson}
\affiliation{Department of Space, Earth and Environment, Chalmers University of Technology, Onsala Space Observatory, SE-439 92 Onsala, Sweden.}
\author[0000-0003-3786-3486]{Seth Redfield}
\affiliation{Astronomy Department and Van Vleck Observatory, Wesleyan University, Middletown, CT 06459, USA}
\author[0000-0003-2058-6662]{George R. Ricker}
\affiliation{MIT Kavli Institute for Astrophysics and Space Research \& MIT Physics Department
}
\author[0000-0002-6892-6948]{Sara Seager}
\affil{Department of Earth, Atmospheric, and Planetary Sciences, Massachusetts Institute of Technology, Cambridge, MA 02139, USA}
\affil{Department of Physics and Kavli Institute for Astrophysics and Space Research, Massachusetts Institute of Technology, Cambridge, MA 02139, USA}
\affil{Department of Aeronautics and Astronautics, Massachusetts Institute of Technology, Cambridge, MA 02139, USA}
\author[0000-0001-6763-6562]{Roland Vanderspek}
\affiliation{Department of Physics and Kavli Institute for Astrophysics and Space Research, Massachusetts Institute of Technology, Cambridge, MA 02139, USA}
\author[0000-0001-5542-8870]{Vincent Van Eylen}
\affiliation{Mullard Space Science Laboratory, University College London, Holmbury St Mary, Dorking, Surrey RH5 6NT, UK}
\author[0000-0002-4265-047X]{Joshua N.\ Winn}
\affiliation{Department of Astrophysical Sciences, Princeton University, Princeton, NJ 08544, USA}

\begin{abstract} 
GJ\,367 is a bright (V\,$\approx$\,10.2) M1\,V star that has been recently found to host a transiting ultra-short period sub-Earth on a 7.7 hr orbit. With the aim of improving the planetary mass and radius and unveiling the inner architecture of the system, we performed an intensive radial velocity follow-up campaign with the \harps spectrograph -- collecting 371 high-precision measurements over a baseline of nearly 3 years -- and combined our Doppler measurements with new \tess observations from sectors 35 and 36. We found that GJ\,367\,b has a mass of $M_\mathrm{b}$\,=\,0.633\,$\pm$\,0.050~M$_{\oplus}$ and a radius of $R_\mathrm{b}$\,=\,0.699\,$\pm$\,0.024~R$_{\oplus}$, corresponding to precisions of 8\% and 3.4\%, respectively. This implies a planetary bulk density of $\rho_\mathrm{b}$\,=\,10.2\,$\pm$\,1.3~g\,cm$^{-3}$, i.e., 85\% higher than Earth's density. We revealed the presence of two additional non transiting low-mass companions with orbital periods of $\sim$11.5 and 34 days and minimum masses of $M_\mathrm{c}\sin{i_\mathrm{c}}$\,=\,4.13\,$\pm$\,0.36~M$_{\oplus}$ and  $M_\mathrm{d}\sin{i_\mathrm{d}}$\,=\,6.03\,$\pm$\,0.49~M$_{\oplus}$, respectively, which lie close to the 3:1 mean motion commensurability. GJ\,367\,b joins the small class of high-density planets, namely the class of super-Mercuries, being the densest ultra-short period small planet known to date. Thanks to our precise mass and radius estimates, we explored the potential internal composition and structure of GJ\,367\,b, and found that it is expected to have an iron core with a mass fraction of 0.91$^{+0.07}_{-0.23}$. How this iron core is formed and how such a high density is reached is still not clear, and we discuss the possible pathways of formation of such a small ultra-dense planet. 
\end{abstract}

\keywords{radial velocity, transit photometry}

\section{Introduction}\label{sec:intro}

Close-in planets with orbital periods of a few days challenge planet formation and evolution theories and play a key role in the architecture of exoplanetary systems \citep{2015Winn_Fabrycky, 2021Zhu}. To date, about 132 ultra-short period (USP) planets, namely planets with orbital periods shorter than 1\,day \citep{2006Sahu,2014SanchisOjeda}, have been validated, and only 36 of these were confirmed and have measured radii and masses\footnote{See \href{https://exoplanetarchive.ipac.caltech.edu}{exoplanetarchive.ipac.caltech.edu}, as of May 2023.}. USPs are preferred targets for transit and radial velocity (RV) planet search surveys, as the transit probability is higher -- it scales as $P^{-2/3}_\mathrm{orb}$ -- and the Doppler reflex motion is larger -- it scales as $P^{-1/3}_\mathrm{orb}$. In addition, their orbital period is typically 1 order of magnitude shorter than the rotation period of the star, allowing one to disentangle bona fide planetary signals from stellar activity \citep{2011Hatzes,2019Hatzes, WINN2020}. 

\cite{2014SanchisOjeda} found that the occurrence rate of rocky USP planets seems to depend on the spectral type of the host star, being 0.15$\,\pm\,$0.05\,\% for F dwarfs, 0.51$\,\pm\,$0.07\,\% for G dwarfs, and 1.10$\,\pm\,$0.40\,\% for M dwarfs. In this context, low-mass stars, such as M dwarfs, are particularly suitable to search for close-in terrestrial planets. Given the relatively small stellar radius and mass, a planet transiting an M dwarf star induces both a deeper transit and a larger RV signal, increasing its detection probability \citep{Cifuentes2020}.

The formation process of USP planets is still not fully understood, and different scenarios have been proposed to explain their short-period orbits: dynamical interactions in multi-planet systems \citep{2010Schlaufman}; low-eccentricity migration due to secular planet-planet interactions \citep{2019Pu}; high-eccentricity migration due to secular dynamical chaos \citep{2019Petrovich}; tidal orbital decay of USP planets formed in situ \citep{2017Lee_Chiang}; and obliquity-driven tidal migration \citep{2020Millholland}. Intensive follow-up observations of systems hosting USP planets can help us to understand the formation and evolution mechanisms of short-period objects and other phenomena related to star$-$planet interactions \citep{Serrano2022}. 

\cite{2014SanchisOjeda}, \cite{2017Adams}, and \cite{WINN2020} found that most USP planets have nearby planetary companions. In multi-planet systems, USP planets show wider-than-usual period ratios with their nearest companion, and appear to have larger mutual inclinations than planets on outer orbits \citep{2018Rodriguez,2018Dai}. These observations suggest that USP planets experienced a change in their orbital parameters, such as inclination increase and orbital shrinkage, suggesting the presence of long-period planets.

During its primary mission, NASA's Transiting Exoplanet Survey Satellite \citep[\tess;][]{Ricker2015} discovered a shallow ($\sim$300\,ppm) transit event repeating every $\sim$7.7\,hr ($\sim$0.32\,days) and associated to a USP small planet candidate orbiting the bright ($V$\,$\approx$\,10.2), nearby (d\,$\approx$\,9.4\,pc), M1\,V star GJ\,367. \cite{2021Lam} recently confirmed GJ\,367\,b as a bona fide USP sub-Earth with a radius of R$_\mathrm{b}$\,=\,0.718\,$\pm$\,0.054 R$_{\oplus}$ and a mass of M$_\mathrm{b}$\,=\,0.546\,$\pm$\,0.078\,M$_\oplus$. 

As the majority of well-characterized USP systems are consistent with having additional planetary companions \citep{2021Dai}, it is quite realistic to believe that the GJ\,367 hosts more than one planet. As part of the RV follow-up program carried out by the KESPRINT consortium\footnote{\url{https://kesprint.science/}.}, we here present the results of an intensive RV campaign conducted with the \harps spectrograph to refine the mass determination of the transiting USP planet and search for external planetary companions, while probing the architecture of the GJ\,367 planetary system.

The paper is organized as follows: we provide a summary of the \tess data and describe our \harps spectroscopic follow-up in Sect.~\ref{sec:Observations}. Stellar fundamental parameters are presented in Sect.~\ref{sec:Star_param}. We report on the RV and transit analysis in Sects.~\ref{sec:Frequency_analysis} and \ref{sec:Analysis}, along with the frequency analysis of our \harps time series. Discussion and conclusions are given in Sects.~\ref{sec:Discussion} and \ref{sec:Conclusions}, respectively.

\section{Observations}\label{sec:Observations}
\subsection{TESS Photometry}\label{tess_obs}

\tess observed GJ\,367 in Sector 9 as part of its primary mission, from 2019 February 28 to 2019 March 26, with CCD\,1 of camera 3 at a cadence of 2 minutes. These observations have been presented in \cite{2021Lam}. About 2 yr later, \tess re-observed GJ\,367 as part of its extended mission in Sectors 35 and 36, from 2021 February 9 to 2021 April 2, with CCD\,1 and 2 of camera 3 at a higher cadence of 20s as well as at 2 minutes. The photometric data were processed by the \tess\ data processing pipeline developed by the Science Processing Operations Center (SPOC; \citealp{SPOC}). The SPOC pipeline uses Simple Aperture Photometry (SAP) to generate stellar light curves, where common instrumental systematics are removed via the Presearch Data Conditioning (PDCSAP) algorithm developed for the \kepler space mission \citep{2012Stumpe,2014Stumpe,2012Smith}.

We retrieved \tess\ Sector 9, 35, and 36 data from  from the Mikulski Archive for Space Telescopes (MAST)\footnote{\url{https://mast.stsci.edu/portal/Mashup/Clients/Mast/Portal.html}.} and performed our data analyses using the PDCSAP light curve. We ran the D\'etection Sp\'ecialis\'ee de Transits  \citep[DST;][]{2012Cabrera} algorithm to search for additional transit signals and found no significant detection besides the 7.7\,h signal associated to GJ\,367\,b, suggesting that there are no other transiting planets in the system observed in \tess Sector 9, 35, and 36, consistent with the SPOC multi-transiting planet search.

\subsection{HARPS high-precision Doppler follow-up}

GJ\,367 was observed with the High Accuracy Radial velocity Planet Searcher (\harps) spectrograph \citep[][]{Mayor2003}, mounted at the ESO-3.6\,m telescope of La Silla Observatory in Chile. We collected 295 high-resolution ($R\,\approx\,115,000$, $\lambda$\,$\in$\,378--691\,nm) spectra between 2020 November 9 and 2022 April 18 (UT), as part of our large observing program 106.21TJ.001 (PI: Gandolfi) to follow-up \tess\ transiting planets. When added to the 77 \harps\ spectra published in \citet{2021Lam}, our data includes 371 \harps\ spectra.

The exposure time varied between 600 and 1200 s, depending on weather conditions and observing schedule constraints, leading to a signal-to-noise ratio (S/N) per pixel at 550\,nm ranging between 20 and 90, with a median of $\sim$55. We used the second fiber of the instrument  to simultaneously observe a Fabry-Perot interferometer and trace possible nightly instrumental drifts \citep{Wildi2010,Wildi2011}. The \harps data were reduced using the Data Reduction Software \citep[DRS;][]{Lovis2007} available at the telescope. The RV measurements, as well as the H$\alpha$, H$\beta$, H$\gamma$, Na D activity indicators, log\,R$'_{HK}$, the differential line width (DLW), and the chromaticity index (CRX), were extracted  using the codes NAIRA \citep{2017Astudillo-Defru} and \serval \citep{2017Zechmeister}. NAIRA and \serval feature template matching algorithms that are suitable to derive precise RVs for M-dwarf stars, when compared to the cross-correlation function technique implemented in the DRS. We tested both the NAIRA and \serval\ RV time series and found no significant difference in the fitted parameters. While we have no reason to prefer one code over the other, we used the RV data extracted with NAIRA for the analyses described in the following sections.

Table~\ref{table-GJ367-RV} lists the \harps RVs, including those previously reported in \cite{2021Lam}, along with the activity indicators and line profile variation diagnostics extracted with NAIRA and \serval. Time stamps are given in Barycentric Julian Date in the Barycentric Dynamical Time (BJD$_{\mathrm{TDB}}$).

\begin{deluxetable*}{lcc}[!ht]
\tablenum{1}
\tablecaption{Fundamental parameters of GJ\,367.\label{tab:1}}
\tablewidth{0pt}
\tablehead{{Parameter} & \colhead{Value} & \colhead{Reference}}
\startdata
Name & GJ\,367 & \\
     & TOI-731 & \\
     & TIC\,34068865 & \\
   R.A. (J2000) & 09:44:29.15 & [1]  \\
   Decl. (J2000) & $-$45:46:44.46 & [1]      \\
   TESS-band magnitude & 8.032 $\pm$ 0.007 & [2] \\
   V-band magnitude & 10.153 $\pm$ 0.044 & [3]     \\
   Parallax (mas) &  106.173 $\pm$ 0.014 & [1]  \\
   Distance (pc) & 9.413 $\pm$ 0.003 &  [1]  \\
   Star mass $M_{*}$ (M$_{\odot}$) & 0.455 $\pm$ 0.011 & [4]  \\
   Star radius $R_{*}$ (R$_{\odot}$) & 0.458 $\pm$ 0.013 & [4] \\
   Effective temperature T$_\mathrm{eff}$ (K) & 3522 $\pm$ 70 & [4] \\
   Stellar density $\rho_{*}$ ($\rho_{\odot}$) & 4.75$_{-0.39}^{+0.44}$ & [4] \\
   Metallicity [Fe/H] & $-$0.01 $\pm$ 0.12 & [4] \\
   Surface gravity $log\,g_\star$ & 4.776 $\pm$ 0.026 & [4] \\
   Luminosity $L_{*}$ ($L_{\odot}$) & 0.0289$_{-0.0027}^{+0.0029}$ & [4] \\
   log\,R$'_{HK}$ & -5.169 $\pm$ 0.068 & [4] \\
   Spectral type & M1.0\,V & [5] \\
\enddata
\tablecomments{[1] \cite{2021GaiaEDR3}, [2] \tess input catalog (TIC; \cite{2018Stassun, 2019Stassun}), [3] \cite{2022Paegert}, [4] This work, [5] \cite{2010MNRAS.403.1949K}.}
\end{deluxetable*}

\subsection{HARPS spectropolarimetric observations}

With the aim of measuring the magnetic field of GJ\,367, we performed a single circular polarization observation with the HARPSpol polarimeter \citep{piskunov:2011,snik:2011} on 2022 November 16 (UT), as part of the our ESO HARPS program 1102.C-0923 (PI: Gandolfi). We used an exposure time of 3600\,s split in four $T_\mathrm{exp}$\,=\,900 s sub-exposures obtained with different configuration of polarization optics to ensure cancellation of the spurious instrumental signals \citep[see][]{donati:1997,bagnulo:2009}. The data reduction was carried out with the REDUCE code \citep{piskunov:2002} following the steps described in \citet{rusomarov:2013}. The resulting Stokes $I$ (intensity) and Stokes $V$ (circular polarization) spectra cover approximately the same wavelength interval as the usual HARPS observations at a slightly reduced resolving power ($R\approx110,000$). We also derived a diagnostic null spectrum \citep[e.g.][]{bagnulo:2009}, which is useful for assessing the presence of instrumental artifacts and non-Gaussian noise in the Stokes $V$ spectra.

\section{Stellar parameters}
\label{sec:Star_param}

\subsection{Photospheric and fundamental parameters}
\label{StarParameters}

We derived the spectroscopic parameters using the new co-added HARPS spectrum for GJ\,367. Following the prescription in \citet{2018AJ....155..127H}, we first estimate the stellar effective temperature $T_\mathrm{eff}$, metallicity [Fe/H], and radius $R_\star$ using {\tt SpecMatch-Emp} \citep{2017ApJ...836...77Y}. The code attempts to find a subset of best-matching template spectra from the library to the input spectrum and derives the best empirical values for the above parameters. {\tt SpecMatch-Emp} returned $T_\mathrm{eff}= 3522 \pm 70$ K, $R_\star= 0.452 \pm 0.045\,R_\odot$, and $\mathrm{[Fe/H]}=-0.01 \pm 0.12$. We then used those parameters to estimate the other stellar parameters as well as refine $R_\star$. As described in \citet{2021AJ....162..161H}, we implemented a Markov Chain Monte Carlo (MCMC) simulation and derived the parameters in a self-consistent manner, making use of the empirical formulae by \citet{2015ApJ...804...64M} and \citet{2019ApJ...871...63M} for the derivations of the stellar mass $M_\star$ and radius $R_\star$. We found that GJ\,367 has a mass of $M_\star$\,=\,0.455\,$\pm$\,0.011\,M$_\odot$ and a radius of $R_\star$\,=\,0.458\,$\pm$\,0.013\,R$_\odot$, with the latter in very good agreement with the value derived using {\tt SpecMatch-Emp}. In the MCMC implementation, we also derived the stellar density $\rho_\star$, surface gravity $\log g_\star$, and luminosity $L_\star$ of the star. Results of our analysis are listed in Table~\ref{tab:1}.
As the quoted uncertainties of the stellar parameters do not account for possible unknown systematic errors -- which in turn might affect the estimates of the planetary parameters -- we performed a sanity check and determined the stellar mass and radius using the code BASTA \citep{2022MNRAS.509.4344A}. We fitted the derived stellar parameters to the BaSTI isochrones \citep[][science case 4 in Table~1 of the BASTA paper]{2018ApJ...856..125H}. Starting from the effective temperature, metallicity, and radius, as derived from {\tt SpecMatch-Emp}, yielded a mass of $M_\star$\,=\,0.435$_{-0.040}^{+0.035}$\,M$_\odot$ and radius of $R_\star$\,=\,0.411$_{-0.034}^{+0.032}$\,R$_\odot$. The stellar mass is in good agreement with the previous result. The new estimate of the stellar radius is smaller than the value reported in Table~\ref{tab:1}, but it is still consistent within $\sim$1.4\,$\sigma$ (where $\sigma$ is the sum in quadrature of the two nominal uncertainties) with a p-value of $\sim$15\%. Assuming a significance level of 5\%, the two radii are consistent, providing evidence that our estimates might not be significantly affected by inaccuracy.
 
\subsection{Rotation period}
\label{sec:star_rot}

Using archival photometry from the Wide Angle Search for Planets survey (WASP), \cite{2021Lam} found a photometric modulation with a period of 48\,$\pm$\,2~days. \cite{2021Lam} also measured a Ca\,{\sc II} H\,\&\,K chromospheric activity index of log\,R$'_\mathrm{HK}$\,=\,$-5.214$\,$\pm$\,0.074 from their 77 \harps spectra. Based on the log\,R$'_\mathrm{HK}-$rotation empirical relationship for M-dwarfs from \cite{Astudillo_Defru_2017}, they estimated a stellar rotation period of $P_\mathrm{rot}$\,=\,58.0\,$\pm$\,6.9~days. 

We independently derived a Ca\,{\sc II} H\,\&\,K chromospheric activity index of log\,R$'_\mathrm{HK}$\,=\,-5.169\,$\pm$\,0.068 from the 371 \harps\ spectra and estimated the rotation period of GJ\,367 using the same empirical relationship. We found a rotation period of $P_\mathrm{rot}$\,=\,54\,$\pm$\,6~d, in good agreement with the previous estimated value. We note that our estimate is consistent within 1$\sigma$ with the value of $P_\mathrm{rot}$\,=\,51.30\,$\pm$\,0.13~d recovered by our sinusoidal signal analysis described in Sect.\,\ref{RV_analyis}, and with the period of $P_\mathrm{rot}\,=\,53.67 _{-0.53}^{+0.65}$~d derived by our multidimensional Gaussian process (GP) analysis described in Sect.\,\ref{GP}.

\subsection{Magnetic field}
\label{spectropolarimetry}

\begin{figure}[t!]
\centering
\includegraphics[trim=0.3cm 3.2cm 0.0cm 2.8cm,clip=false, width=\linewidth]{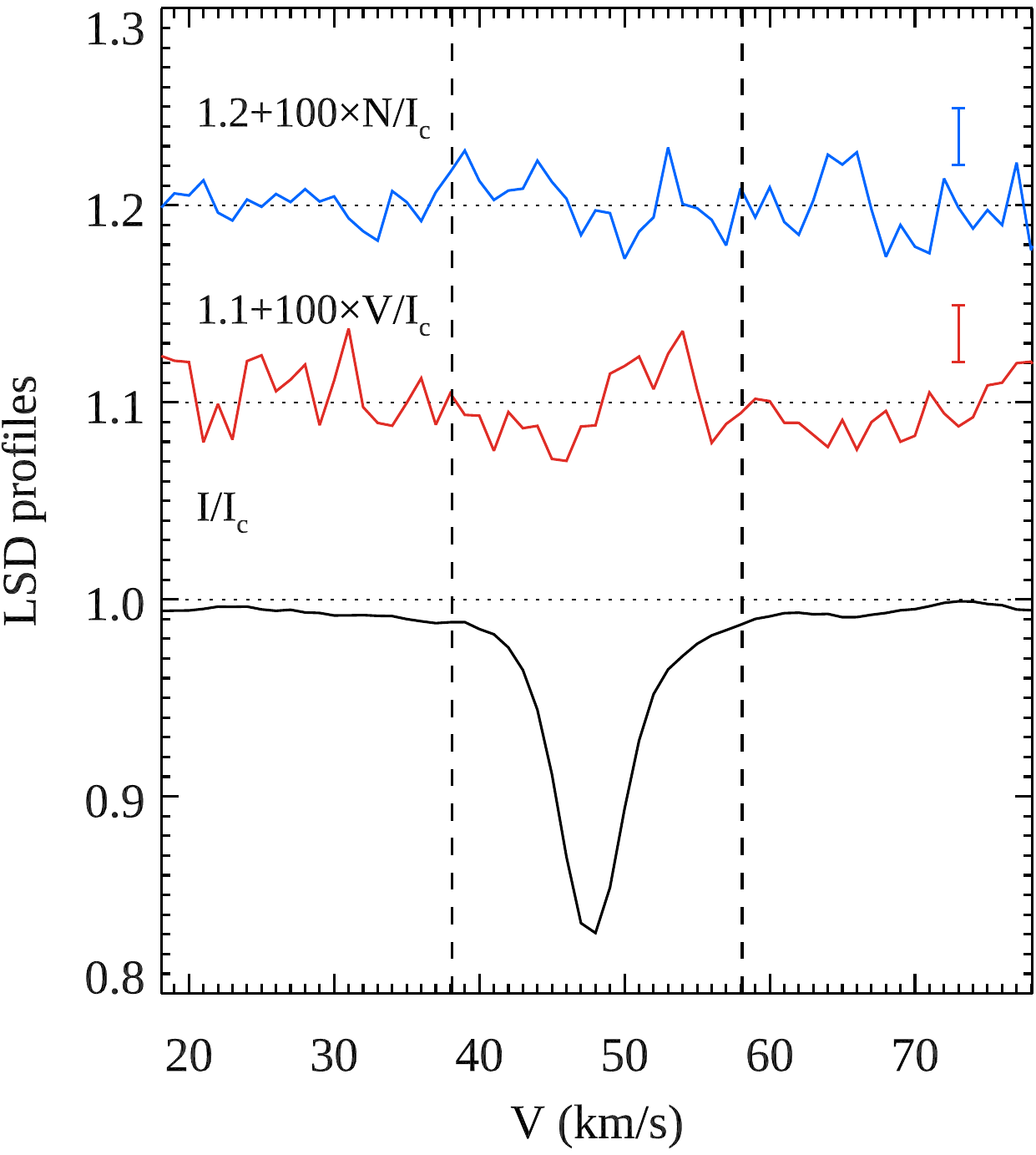}
    \caption{Least-squares deconvolved (LSD) Stokes $I$, $V$ and null profiles of GJ~367. The polarization profiles are shifted vertically and expanded by a factor of 100 relative to the intensity profile. The vertical dashed lines indicate velocity interval adopted for the longitudinal field measurement.
\label{fig:LSDprofile}}
\end{figure}

Our spectropolarimetric observation of GJ\,367 achieved a median S/N of about 90 over the red HARPS chip. This is insufficient for detecting Zeeman polarization signatures in individual lines even for the most active M dwarfs. To boost the signal, we made use of the least-squares deconvolution procedure \citep[LSD,][]{donati:1997} as implemented by \citet{kochukhov:2010}. The line mask required for LSD was obtained from the VALD database \citep{ryabchikova:2015} using the atmospheric parameters of GJ\,367 and assuming solar abundances (Sect.~\ref{StarParameters}). We used about 5000 lines deeper than 20\% of the continuum for LSD, reaching a S/N of 7250 per 1~km\,s$^{-1}$ velocity bin.  
The resulting Stokes V profile has a shape compatible with a Zeeman polarization signature with an amplitude of $\approx$0.04\% (Figure~\ref{fig:LSDprofile}). However, with a false-alarm probability (FAP) =\,2.3\%, detection of this signal is not statistically significant according to the usual detection criteria employed in high-resolution spectropolarimetry \citep{donati:1997}.
The mean longitudinal magnetic field, which represents the disk-averaged line-of-sight component of the global magnetic field, derived from this Stokes $V$ profile is $\langle B_{\rm z} \rangle=-7.3\pm3.2$~G.

Our spectropolarimetric observation of GJ~367 suggests that this object is not an active M dwarf. Its longitudinal magnetic field was found to be below 10~G, which is much weaker than $\ge100$--700~G longitudinal fields typical of active M dwarfs \citep{donati:2008,morin:2008}. Considering that the strength and topology of the global magnetic fields of M dwarfs is systematically changing with the stellar mass \citep{kochukhov:2021}, it is appropriate to compare GJ~367 with early-M dwarfs. To this end, the well-known $\sim$20~Myr old M1V star AU Mic was observed with HARPSpol in the same configuration and with a similar S/N as our GJ~367 observations \citep{kochukhov:2020}, yielding consistent detections of polarization signatures and longitudinal fields of up to 50~G. The magnetic activity of GJ~367 is evidently well below that of AU Mic.

\subsection{Age}
\label{sec:age}

Determining the age of M dwarf stars is especially challenging and depends on the methods being utilized. \cite{2021Lam} estimated an isochronal age of 8.0$^{+3.8}_{-4.6}$\,Gyr and a  gyrochronological age of 4.0\,$\pm$\,1.1\,Gyr for GJ\,367. More recently, \cite{2022MNRAS.513..661B} gave two different estimates: (1) by comparing Gaia EDR3 parallax and photometric measurements with theoretical isochrones, they suggested a young age $<$\,60~Myr. However, as pointed out by the authors, this is not in line with the star’s Galactic kinematics that exclude membership to any nearby young moving group; (2) by considering the Galactic dynamical evolution, which indicates an age of 1–8\,Gyr.

\begin{figure}[t!]
\centering
\includegraphics[width=\linewidth]{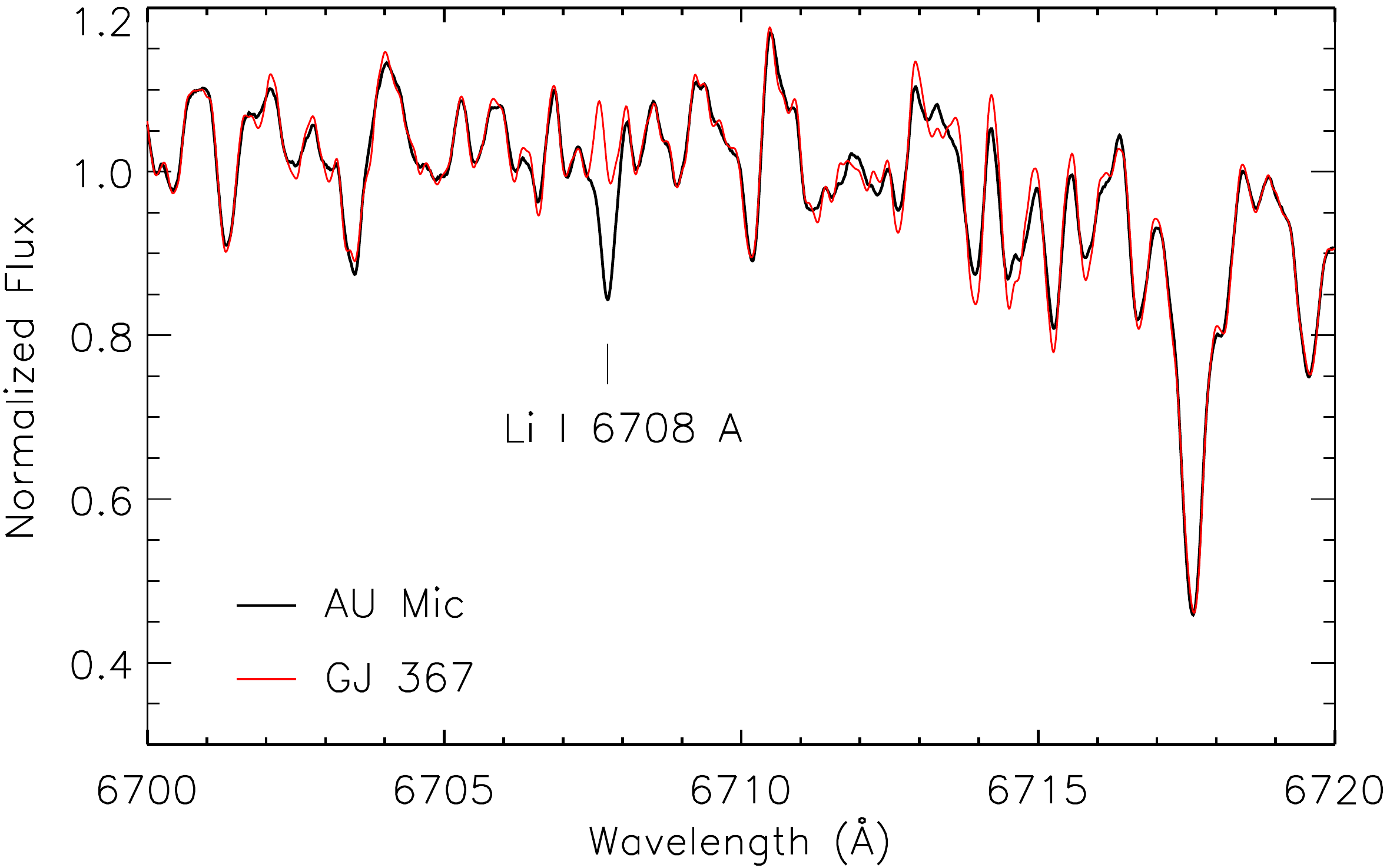}
    \caption{Co-added \harps spectrum of AU\,Mic (black line) in the spectral region encompassing the Li\,{\sc i}\,6708\,\AA \,absorption line. Superimposed with a thick red line is the co-added \harps spectrum of GJ\,367, which has been rotationally broadened to match the $v$\,sin\,$i_\star$\,=\,7.8\,km\,s$^{-1}$ of AU\,Mic.
\label{fig:Lithium}}
\end{figure}

In this respect, the results presented in our study shows compelling evidence that GJ\,367 is not a young star:
\begin{itemize}
\item The time series of our \harps RVs and activity indicators give a clear detection of a spot-induced rotation modulation with a period of about 52$-$54\,d, which translates into a gyrochronological age of $\sim$4.6-4.8\,Gyr \citep{2010BarnesA,2010BarnesB}.
\item The \harps spectra of GJ\,367 show no significant lithium absorption line at 6708\,\AA. Figure~\ref{fig:Lithium} displays the co-added \harps\ spectrum of AU~Mic \citep{2022Zicher,Klein2022} in the spectral region around the Li\,{\sc i}\,6708\,\AA\ line. AU\,Mic is a well-known 22-Myr-old M1 star located in the $\beta$ Pictoris moving group \citep{Mamajek2014}. Superimposed with a thick red line is GJ\,367's co-added \harps\ spectrum, which has been broadened to match the projected rotational velocity of AU~Mic ($v$\,sin\,$i_\star$\,=\,7.8\,km\,s$^{-1}$). Since lithium is quickly depleted in young GKM stars, the lack of lithium in the spectrum of GJ\,367 suggests an age $\gtrsim$\,50\,Myr \citep{2014Binks,2021Binks}.
\item  The low level of magnetic activity inferred by the Ca\,H\,\&\,K indicator log\,R$'_\mathrm{HK}$ (Sect,~\ref{sec:star_rot}) and the weak magnetic field (Sect.~\ref{spectropolarimetry}) are consistent with an old, inactive M-dwarf scenario \citep{Pace2013}.
\item We measured an average H${\alpha}$ equivalent width of EW\,=\,0.0638\,$\pm$\,0.0014\,\AA. Using the empirical relation that connects the H${\alpha}$ equivalent width with stellar age \citep{2021AJ....161..277K}, this translates into an age $\gtrsim$\,300\,Myr. 
\end{itemize}

We therefore conclude that GJ\,367 is a rather slowly rotating, old star with a low magnetic activity level, rather than a young M dwarf. This conclusion is consistent with a recent study by \cite{2023MNRAS.tmp..380G}, who measured an age of 7.95\,$\pm$\,1.31\,Gyr from the M dwarf rotation$-$age relation. 

\section{Frequency analysis of the \harps Time Series}
\label{sec:Frequency_analysis}

In order to search for the Doppler reflex motion induced by the USP planet GJ\,367\,b and unveil the presence of potential additional signals associated with other orbiting companions and/or stellar activity, we performed a frequency analysis of the \harps RV measurements and activity indicators. For this analysis we did not include the \harps measurements\footnote{Twenty-four measurements taken with the old fiber bundle between 2003 December 12 and 2010 February 7 (UT) and 77 measurements acquired between 2019 June 23 and 2019 March 23 (UT) with the new fiber bundle.} presented in \citet{2021Lam} and used only the Doppler data collected between 2020 November 9 and 2022 April 18 (UT) as part of our \harps\ large program, to avoid spurious peaks introduced by the poor sampling of the existing old data set, and to avoid having to account for the RV offset caused by the refurbishment of the instrument.

\begin{figure*}[ht!]
\begin{center}
\includegraphics[clip, trim=0.9cm 2.7cm 0.0cm 4.1cm, width=0.6\textwidth]{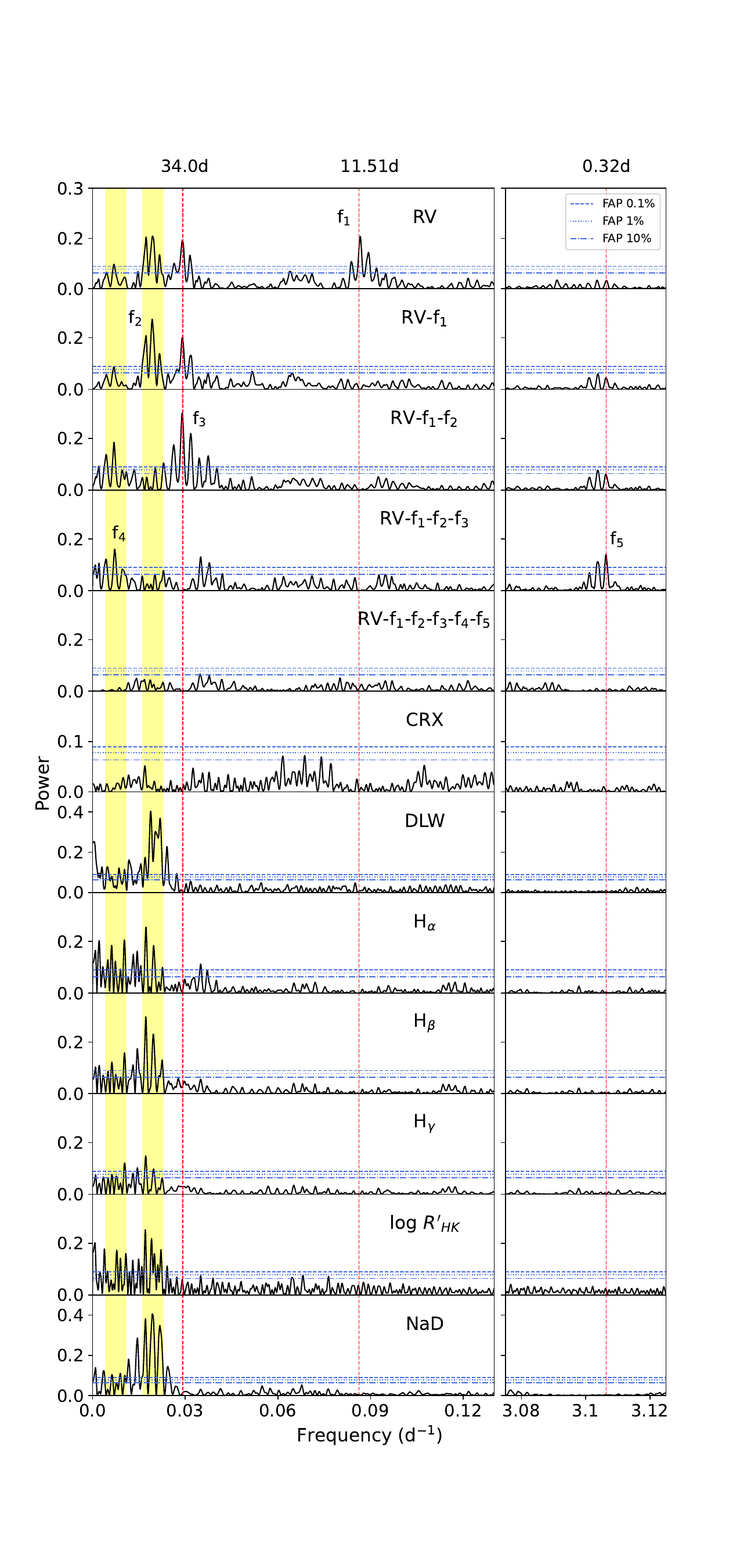}
\caption{Generalized Lomb-Scargle (GLS) periodograms of the \harps RV measurements (\textit{upper panel}); RV residuals after subtracting the $f_{1}$ signal at 11.5 days (\textit{second  panel}), the $f_{2}$ signal related to stellar activity at 52.2 days (\textit{third panel}), the $f_{3}$ signal at 34 days (\textit{fourth panel}), the $f_{4}$ and $f_{5}$=$f_\mathrm{b}$ signals at 115 days and 0.322 days (\textit{fifth panel}). Also shown are periodograms of the activity indicators (\textit{remaining panels}). The 10\%, 1\% and 0.1\% false alarm probabilities (FAPs) estimated using the bootstrap method are shown with horizontal blue lines. The red vertical lines mark the orbital frequencies of the transiting planet GJ\,367\,b ($f_{5}$\,=\,$f_\mathrm{b}$\,=\,3.106\,day$^{-1}$), and of the additional Doppler signals we found in the \harps data, which are  associated to the presence of two additional orbiting planets ($f_1$\,=\,$f_\mathrm{c}$\,=\,0.086 \,day$^{-1}$ and $f_3$\,=   \,$f_\mathrm{d}$\,=\,0.029\,day$^{-1}$). The shaded yellow bands indicate the rotation period of the star centered around f$_2$, and the long-period stellar signal $f_4$.\label{fig:GLS_GJ367}}
\end{center}
\end{figure*}

Figure~\ref{fig:GLS_GJ367}  shows the generalized Lomb$-$Scargle (GLS; \citealp{2009Zechmeister}) periodograms of the \harps RVs and activity indicators in two frequency ranges, i.e. 0.000\,$-$\,0.130\,d$^{-1}$ (left panels) and 3.075\,$-$\,3.125\,d$^{-1}$ (right panels), with the former including the frequencies at which we expect to see activity signals at the rotation period of the star, and the latter encompassing the orbital frequency of the transiting planet GJ\,367\,b. The horizontal dashed lines mark the GLS powers at the 0.1\%, 1\%, and 10\% FAP. The FAP was estimated following the bootstrap method described in \cite{murdoch_1993}, i.e., by computing the GLS periodogram of 10$^6$ mock time series obtained by randomly shuffling the measurements and their uncertainties, while keeping the time stamps fixed. In this work we assumed a peak to be significant if its FAP\,$<$\,0.1\%.  

The GLS periodogram of the \harps RVs (Fig.~\ref{fig:GLS_GJ367}, upper panel) shows its most significant peak at $f_1$\,=\,0.086\,day$^{-1}$, corresponding to a period of about 11.5~days. This peak is not detected in the activity indicators, providing evidence that the 11.5 day signal is caused by a second planet orbiting the host star, hereafter referred to as GJ\,367\,c.

We used the pre-whitening technique \citep{2010Hatzes} to identify additional significant signals and successively remove them from the RV time series. We employed the code \pyaneti \citep{2018Barragan,2022Barragan} to subtract the 11.5 day signal from the \harps RVs assuming a circular model, adopting uniform priors centered around the period and phase derived from the periodogram analysis, while allowing the systemic velocity and RV semi-amplitude to uniformly vary over a wide range. 

The periodogram of the RV residuals following the subtraction of the signal at $f_1$ (Fig.~\ref{fig:GLS_GJ367}, second panel) shows its most significant peak at $f_2$\,=\,0.019~day$^{-1}$ (52.2~days).  Iterating the pre-whitening procedure and removing the signal at $f_2$, we found a significant peak at  $f_3$\,=\,0.029 day$^{-1}$ (34~days). This peak has no significant counterpart in the activity indicators, suggesting it is associated to a third planet orbiting the star, hereafter referred as GJ\,367\,d (Fig.~\ref{fig:GLS_GJ367}, third panel). The periodograms of the CRX, DLW, H$\alpha$, H$\beta$, H$\gamma$, $\log\,R'_\mathrm{HK}$, and Na\,D show significant peaks in the range 48$-$54~days (Fig.~\ref{fig:GLS_GJ367}, lower panels), i.e., close to the rotation period of the star, suggesting that the peak at $f_2$\,=\,0.019~day$^{-1}$ (52.2~days) seen in the RV residuals is very likely associated to the presence of active regions appearing and disappearing on the visible stellar hemisphere as the star rotates about its axis.

After removing the signal at $f_3$, we found significant power at $f_4$\,=\,0.009\,day$^{-1}$ ($\sim$115\,days; Fig.~\ref{fig:GLS_GJ367}, fourth panel). The periodograms of the activity indicators show also significant power around $f_4$, providing evidence this signal is associated to stellar activity. As we will discuss in Sect.~\ref{GP}, we interpreted the power at $f_4$ as the evolution timescale of active regions. Finally, we found that the RV residuals also show a significant fifth peak at $f_5$\,=\,$f_\mathrm{b}$\,=\,3.106\,day$^{-1}$ (0.322~days), the orbital frequency of the USP planet GJ\,367\,b, further confirming the planetary nature of the transit signal identified in the \tess\ data and announced by \citet{2021Lam}.

Finally, we computed the GLS periodogram of the \harps RVs including all of the available data, i.e., the data acquired before and after the refurbishment of the instrument. Adding the old data points increases the baseline of our observations and, consequently, the frequency resolution.  However, the resulting periodogram is ``jagged'', owing to the presence of aliases with very small frequency spacing, making it more difficult to identify the true peaks.

\section{Data Analysis}
\label{sec:Analysis}

We modeled the \tess transit light curves and \harps RV measurements using three different approaches. The methods differ in the way the Doppler data are fitted, as described in the following subsections.

\subsection{Floating chunk offset method}
\label{FCO}

We used the floating chunk offset (FCO) method to determine the semi-amplitude $K_\mathrm{b}$ of the Doppler reflex motion induced by the USP planet GJ\,367\,b. Pioneered by \citet{2011Hatzes} for the mass determination of the USP planet CoRoT-7\,b, the FCO method relies on the reasonable assumption that, within a single night, the RV variation of the star is mainly induced by the orbital motion of the USP planet rather than stellar rotation, magnetic cycles, or orbiting companions on longer-period orbits. As the RV component due to long-period phenomena can thus be treated as constant within a given night, introducing nightly offsets filters out any long-term RV variations, allowing one to disentangle the reflex motion of the USP planet from additional long-period Doppler signals.

With an orbital period of only 7.7\,hr, the USP planet GJ\,367\,b is suitable to the FCO method \citep[see, e.g.,][]{Gandolfi_2017, 2018Barragan_K2141}. GJ\,367 is accessible for up to 8\,hr at an airmass\,$<$\,1.5 (i.e, altitude\,$>$\,40$^\circ$) from La Silla Observatory, allowing one to cover one full orbit in one single night by acquiring multiple \harps spectra per night. Within the nightly visibility window of GJ\,367, the phase of the long-term signals does not change significantly, the variation being 0.029, 0.010, and 0.006 for the 11.5, 34, and 52 day signals, respectively.

We simultaneously modeled the \tess\ transit light curves and \harps\ RV measurements using the open source software suite \texttt{pyaneti} \citep{2018Barragan,2022Barragan}. The code utilizes a Bayesian approach in combination with MCMC sampling to infer the parameters of planetary systems. The photometric data included in the analysis are subsets of the \tess\ light curve. We selected 2.5 hr of \tess\ data points centered around each transit and detrended each light curve segment by fitting a second-order polynomial to the out-of-transit data. Following \citet{2011Hatzes}, we divided the \harps\ RVs into subsets (``chunks'') of nightly measurements and analyzed only those chunks containing at least two RVs per observing night, leading to a total of 96 chunks.

\begin{figure}[t!]
\centering
\includegraphics[width=\linewidth]{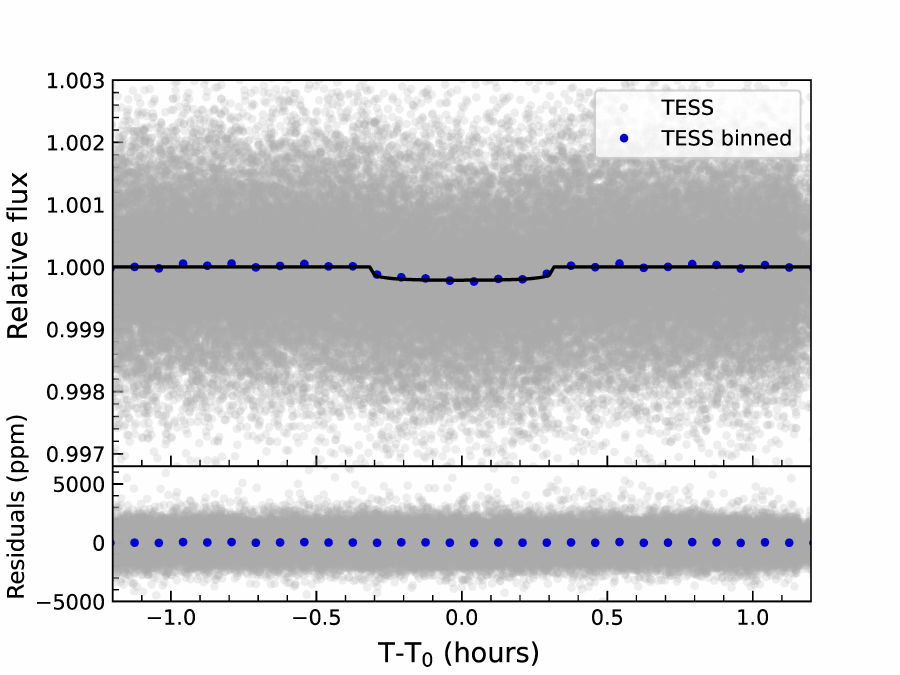}
    \caption{\tess\ transit light curve of GJ\,367\,b and best-fitting model folded at the orbital period of the planet.
    \label{TESS_transit}}
\end{figure}

\begin{figure}[t!]
\centering
\includegraphics[width=\linewidth]{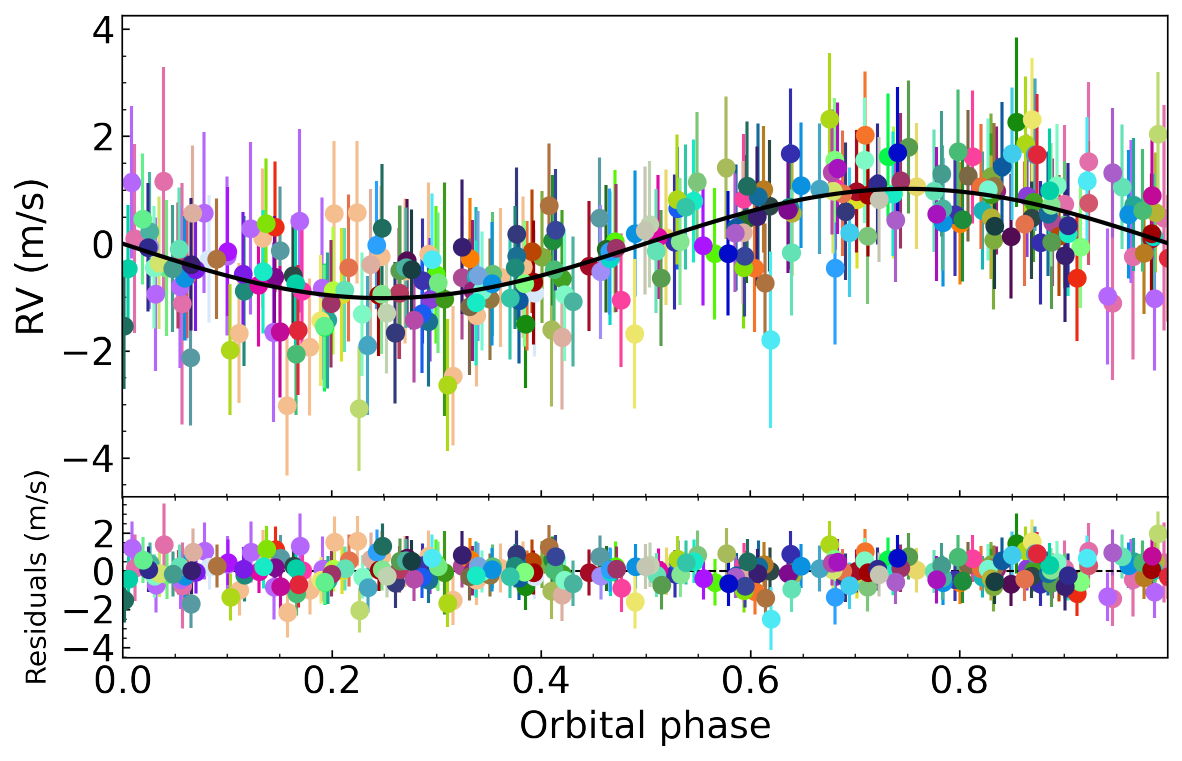}
\caption{\harps\ RVs of GJ\,367 phase-folded at the orbital period of the USP planet and best-fitting model as derived using the FCO method. The different colors refer to the 96 different nightly chunks, which include at least two measurements per night.
\label{RVcurve_FCO}}
\end{figure}

The RV model includes one Keplerian orbit for the transiting planet GJ\,367\,b and 96 nightly offsets. We fitted for a non zero eccentricity adopting the  parameterization proposed by \cite{Anderson_2010} for the eccentricity $e$ and the argument of periastron of the stellar orbit $\omega_\star$ (i.e., $\sqrt{e}\sin{\omega_\star}$ and $\sqrt{e}\cos{\omega_\star}$). We fitted for a photometric and an RV jitter term to account for any instrumental noise not included in the nominal \tess\ and \harps\ uncertainties. We used the limb-darkened quadratic model by \cite{2002Mandel&Agol} for the transit light curve. We adopted Gaussian priors for the limb darkening coefficients, using the values derived by \cite{2017Claret} for the \tess passband, and we imposed conservative error bars of 0.1 on both the linear and the quadratic limb-darkening term. As the shallow transit light curve of GJ\,367\,b poorly constrains the scaled semi-major axis ($a_\mathrm{b}/R_\star$), 
we sampled for the stellar density $\rho_{\star}$ using a Gaussian prior on the star's mass and radius as derived in Sect.~\ref{sec:Star_param}, and recovered the scaled semi-major axis of the planet using the orbital period and Kepler’s third law of planetary motion \citep[see, e.g.][]{2010Winn}. We adopted uniform priors over a wide range for all of the remaining model parameters.
We ran 500 independent Markov chains.  
The posterior distributions were created using the last 5000 iterations of the converged chains with a thin factor of 10, leading to a distribution of 250,000 data points for each model parameter. The chains were initialized at random values within the priors ranges. This ensured a homogeneous sampling of the parameter space.
We followed the same procedure and convergence test as described in \cite{2018Barragan}.
The final estimates and their 1$\sigma$ uncertainties were taken as the median and 68\% of the credible interval of the posterior distributions. Table~\ref{table:2} reports prior ranges and  posterior values of the fitted and derived system parameters. Figure~\ref{RVcurve_FCO} displays the phase-folded RV curve with our \harps data, along with the best-fitting Keplerian model. Different colors refer to different nights. Figure~\ref{TESS_transit} shows the phase-folded transit light curve of GJ\,367\,b, along with the \tess data and best-fitting transit model. We found an RV semi-amplitude variation of $K_\mathrm{b}$\,=\,1.003\,$\pm$\,0.078~m\,s$^{-1}$, which translates into a planetary mass of $M_\mathrm{b}$\,=\,0.633\,$\pm$\,0.050~M$_\oplus$ (7.9\,\% precision). The depth of the transit light curve implies a radius of $R_\mathrm{b}$\,=\,0.699\,$\pm$\,0.024 R$_\oplus$ (3.4\% precision) for GJ\,367\,b. When combined together, the planetary mass and radius yield a mean density of $\rho_\mathrm{b}$\,=\,10.2\,$\pm$\,1.3\,g\,cm$^{-3}$ (12.7\% precision). 
The eccentricity of the USP planet ($e_\mathrm{b}\,=\,0.06_{-0.04}^{+0.07}$) is consistent with zero, as expected given the short tidal evolution time-scale and the age of the system. Assuming a circular orbit, our fit gives an RV semi-amplitude of $K_\mathrm{b}$\,=\,1.001\,$\pm$\,0.077~m\,s$^{-1}$, in excellent agreement with the values listed in Table~\ref{table:2}.

We note that 27 of the HARPS RV measurements were taken during transits of GJ\,367\,b. Assuming the star is seen equator-on ($i_\star$\,=\,90°), its rotation period of $P_\mathrm{rot}$\,$\approx$\,51$-$54\,days (Sect.~\ref{sec:star_rot}) implies an equatorial rotational velocity of $v_\mathrm{rot}$\,$\approx$\,\,0.45\,km\,s$^{-1}$. Using the equations in \cite{2018haex.bookE...2T}, we estimated the semi-amplitude of the Rossiter-McLaughlin effect to be $\approx$\,0.05 m\,s$^{-1}$, which is too small to cause any detectable effect given our RV uncertainties.

\begin{deluxetable*}{lcc}
\tablenum{2}
\tablecaption{GJ\,367\,b parameters from the joint FCO and transit modeling with \pyaneti. \label{table:2}}
\tablewidth{0pt}
\tablehead{{GJ\,367\,b} & \colhead{Prior} & \colhead{Derived value}}
\startdata
\textbf{\textit{Model parameters}} & & \\
Orbital period $P_\mathrm{orb, b}$ [days] & $\mathcal{U}$[0.3219221,0.3219229]  & 0.3219225 $\pm$ 0.0000002 \\ 
Transit epoch $T_\mathrm{0,b}$ [BJD$_\mathrm{TDB}-$2,450,000] &  $\mathcal{U}$[8544.13235,8544.14035]  & 8544.13635 $\pm$ 0.00040\\ 
Planet-to-star radius ratio $R_\mathrm{b}/R_\star$ &  $\mathcal{U}$[0.001,0.025] & 0.01399 $\pm$ 0.00028 \\
Impact parameter $b_\mathrm{b}$ & $\mathcal{U}$[0,1] & 0.584$^{+0.034}_{-0.037}$ \\
$\sqrt{e_\mathrm{b}}\sin{\omega_\mathrm{\star,b}}$ & $\mathcal{U}$[-1.0,1.0]  & 0.16$^{+0.17}_{-0.22}$ \\
$\sqrt{e_\mathrm{b}}\cos{\omega_\mathrm{\star,b}}$ & $\mathcal{U}$[-1.0,1.0]  & 0.04 $\pm$ 0.14 \\
Radial velocity semi-amplitude variation $K_\mathrm{b}$ [m\,s$^{-1}$] & $\mathcal{U}$[0,50]   & 1.003 $\pm$ 0.078 \\
\noalign{\smallskip}
\hline
\textbf{\textit{Derived parameters}} & &  \\
Planet mass $M_\mathrm{b}$\,[$M_\oplus$] & --  & 0.633 $\pm$ 0.050 \\
Planet radius $R_\mathrm{b}$\,[$R_\oplus$] & --  & 0.699 $\pm$ 0.024 \\
Planet mean density $\rho_\mathrm{b}$ [g\,cm$^{-3}$] & --  & 10.2 $\pm$ 1.3 \\
Semi-major axis of the planetary orbit $a_\mathrm{b}$ [AU] & --  & 0.00709 $\pm$ 0.00027  \\
Orbit eccentricity $e_\mathrm{b}$ & --  & 0.06$^{+0.07}_{-0.04}$  \\
Argument of periastron of stellar orbit $\omega_{\star,\mathrm{b}}$ [deg] & --  & 66$^{+41}_{-108}$  \\
Orbit inclination $i_\mathrm{b}$ [deg] & --  & 79.89$^{+0.87}_{-0.85}$  \\
Transit duration $\tau_{14,\mathrm{b}}$ [hr] & -- & 0.629 $\pm$ 0.008  \\
Equilibrium temperature T$_\mathrm{eq,b}$ [K] $^{(a)}$ & --  & 1365 $\pm$ 32 \\
Received irradiance F$_\mathrm{b}$ [F$_{\oplus}$] & -- & 579$^{+57}_{-52}$ \\
\noalign{\smallskip}
\hline
\textbf{\textit{Additional model parameters}} & &  \\
Stellar density $\rho_{\star}$ [g\,cm$^{-3}$]& $\mathcal{N}$[6.68,0.59] & 6.76 $\pm$ 0.59 \\
Parameterized limb-darkening coefficient $q_{1}$ & $\mathcal{N}$[0.3766,0.1000]  & 0.343 $\pm$ 0.095  \\
Parameterized limb-darkening coefficient $q_{2}$ & $\mathcal{N}$[0.1596,0.1000]  &  0.163$^{+0.096}_{-0.088}$  \\
Radial velocity jitter term $\sigma_{RV,\harps}$ [m\,s$^{-1}$] & $\mathcal{J}$[0,100]  & 0.43 $\pm$ 0.08  \\
TESS jitter term $\sigma_{\tess}$ & $\mathcal{J}$[0,100]  & 0.00004 $\pm$ 0.00003  \\
\noalign{\smallskip}
\enddata
\tablecomments{$\mathcal{U}[a,b]$ refers to uniform priors between $a$ and $b$; $\mathcal{N}[a,b]$ refers to Gaussian priors with mean $a$ and standard deviation $b$; $\mathcal{J}[a,b]$ refers to Jeffrey's priors between $a$ and $b$. Inferred parameters and uncertainties are defined as the median and the 68.3\,\% credible interval of their posterior distributions.\\
$^{a}$ Assuming zero albedo and uniform redistribution of heat}
\end{deluxetable*}

\subsection{Sinusoidal activity signal modeling}
\label{RV_analyis}

Using the FCO method, we cannot determine the semi-amplitude of the Doppler signals induced by the two outer planets and by stellar activity (Sect.~\ref{sec:Frequency_analysis}). In the analysis described in this section, we treated the RV signals associated with stellar activity as coherent sinusoidal signals.
Once again, we used the code \texttt{pyaneti} and performed an MCMC analysis similar to the one described in Sect.\,\ref{FCO}. The RV model includes three Keplerians, to account for the Doppler reflex motion induced by the three planets GJ\,367 b, c, and d, and two additional sine functions, to account for the activity-induced signals at the rotation period of the star ($\sim$52\,days) and at the evolution timescale of active regions ($\sim$115\,days), as described in Sect.\,\ref{sec:Frequency_analysis}. We used Gaussian priors for the orbital period and time of first transit of GJ\,367\,b as derived in Sect.\,\ref{FCO}, and uniform wide priors for all of the remaining parameters. 
We fitted for a RV jitter term, as well as for a non zero eccentricity both for the USP planet, and for the two outer companions, following the $e$-$\omega_\star$ parameterization proposed by \cite{Anderson_2010}. Details of the fitted parameters and prior ranges are given in Table~\ref{tab:fit_param}. We used 500 independent Markov chains initialized randomly inside the prior ranges. Once all chains converged, we used the last 5000 iterations and saved the chain states every 10 iterations. This approach generated a posterior distribution of 250,000 points for each fitted parameter. 

\begin{figure*}[t!]
\centering
\includegraphics[width = \linewidth]{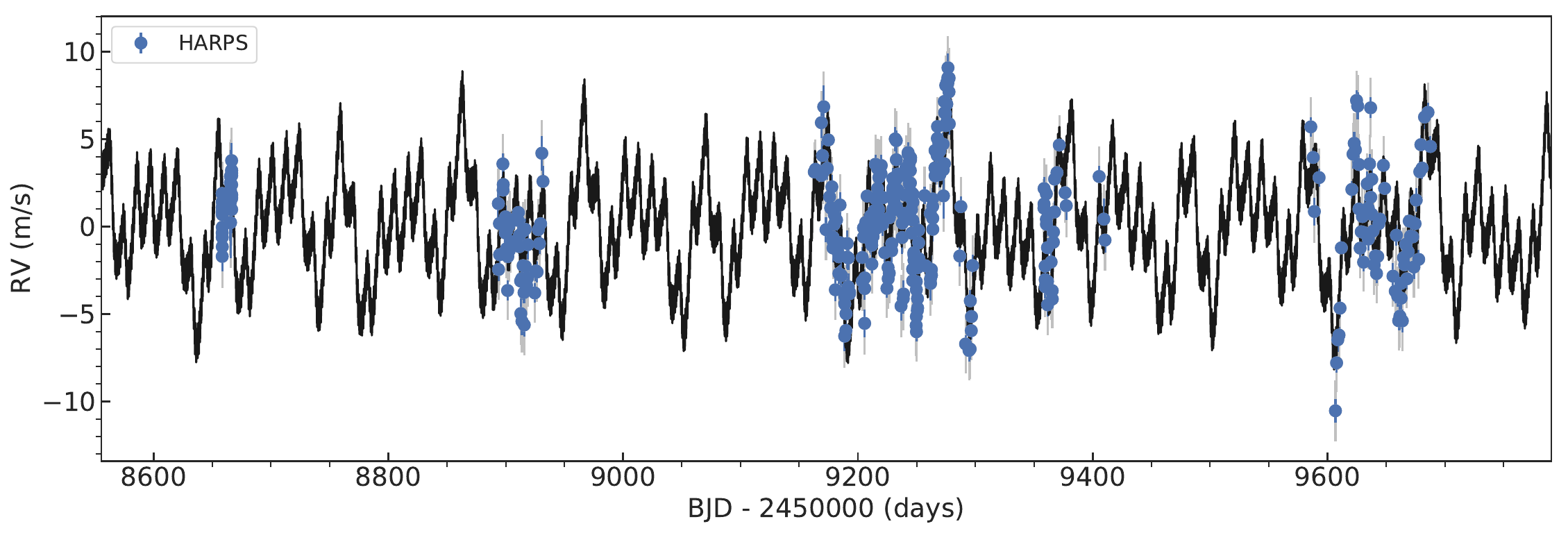}
\caption{\harps RV time series of GJ\,367 along with the best-fitting five-signal model (three planets + stellar rotation + long-period stellar signal). Data are shown as blue filled circles with their nominal uncertainties. The vertical gray lines mark the error bars including the RV jitter.\label{fig:GJ367_rv_fit}}
\end{figure*}

\begin{figure*}[t!]
\centering
\gridline{\fig{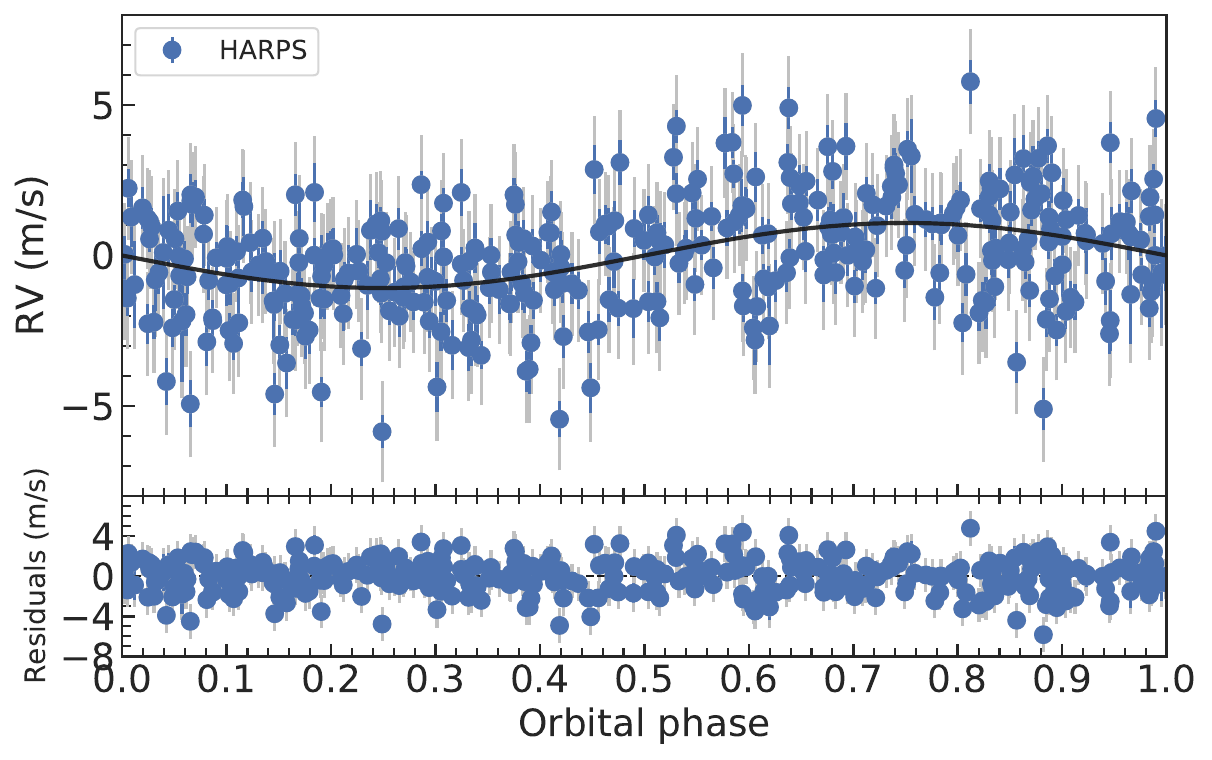}{0.45\textwidth}{(a)}
          \fig{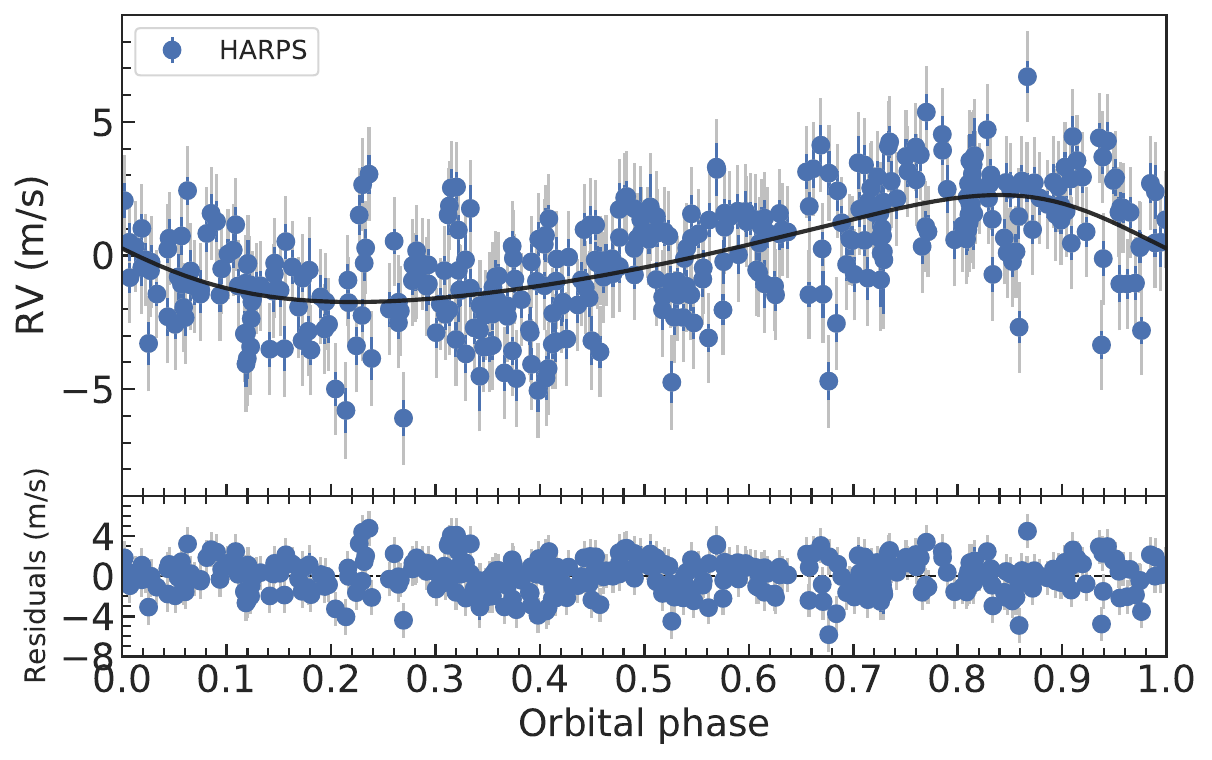}{0.45\textwidth}{(b)}
          }
\gridline{\fig{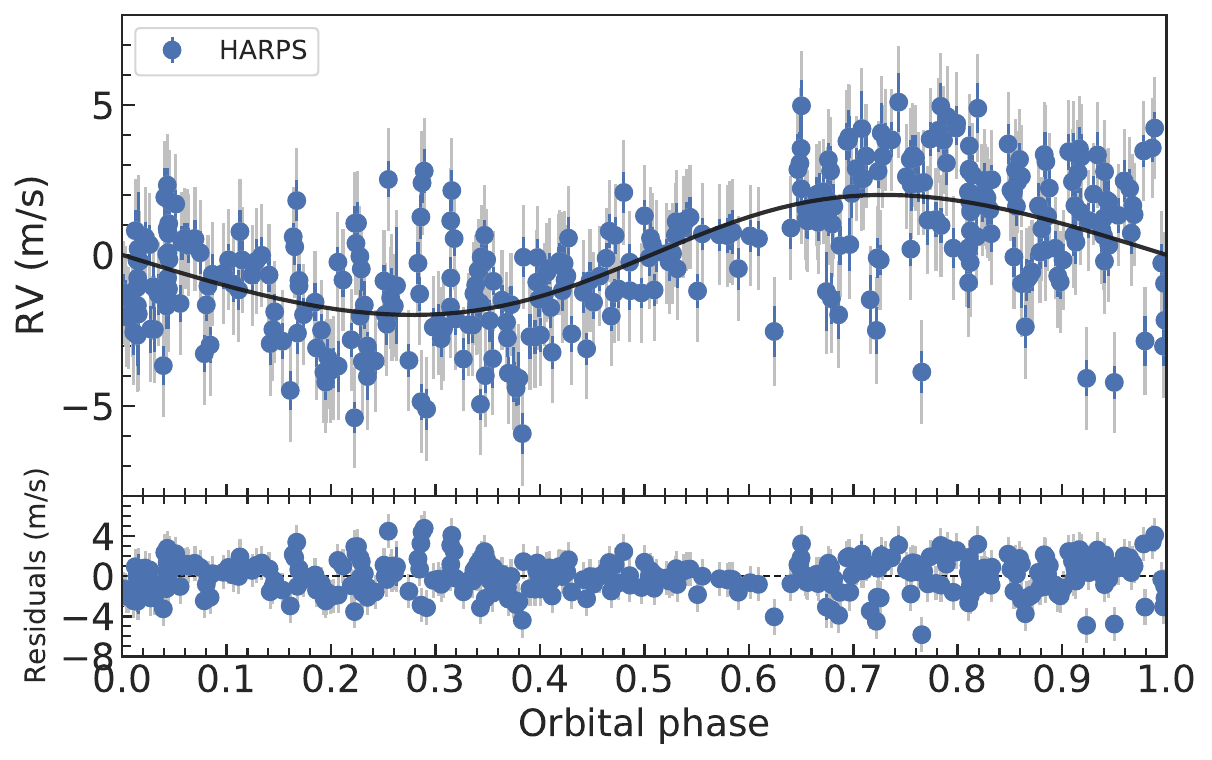}{0.45\textwidth}{(c)}
          \fig{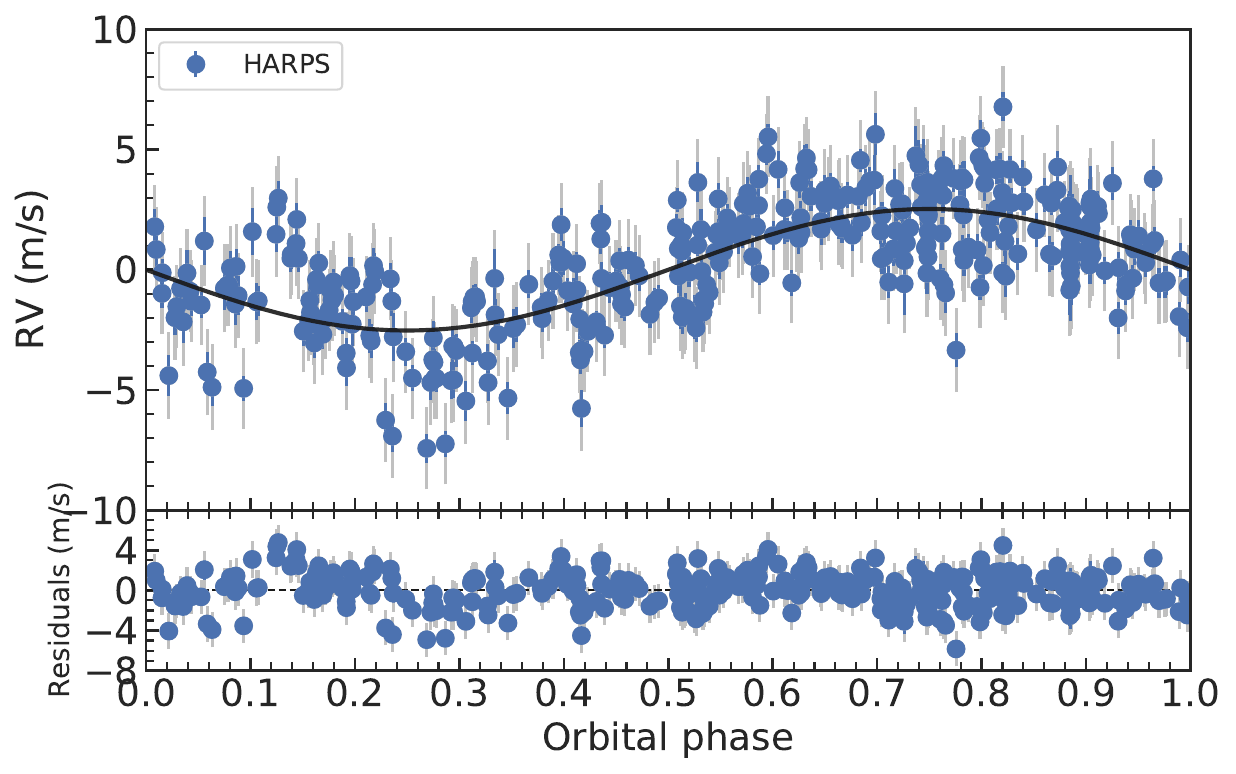}{0.45\textwidth}{(d)}}
\gridline{\fig{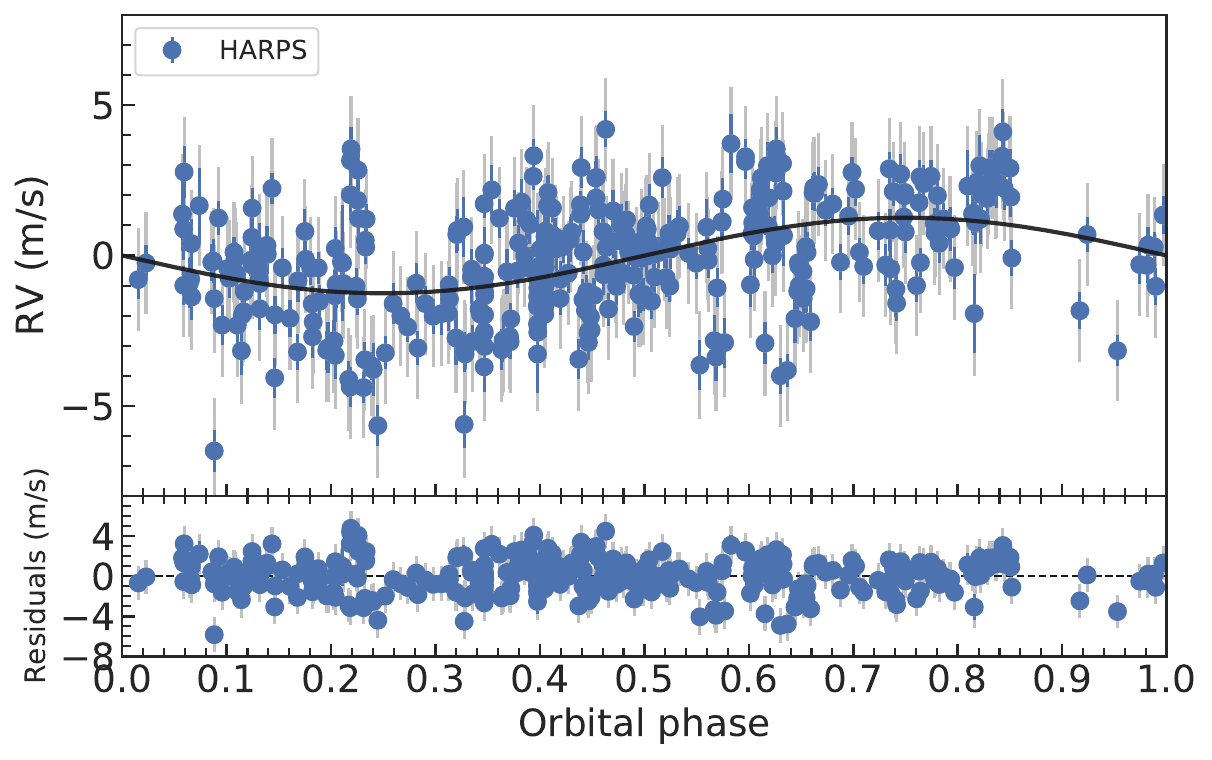}{0.45\textwidth}{(e)}}
\caption{Phase-folded RV curves of GJ\,367\,b (a), GJ\,367\,c (b), GJ\,367\,d (c), stellar rotation (d), and long-period stellar signal (e). Data are shown as blue filled circles with their nominal uncertainties. The vertical gray lines mark the error bars including the RV jitter.\label{fig:GJ367_rv_fit_folded}}
\end{figure*}

The RV semi-amplitude variations induced by the three planets are $K_\mathrm{b}$\,=\,1.10\,$\pm$\,0.14\,m\,s$^{-1}$, $K_\mathrm{c}$\,=\,2.01\,$\pm$\,0.15\,m\,s$^{-1}$, and $K_\mathrm{d}$\,=\,1.98\,$\pm$\,0.15\,m\,s$^{-1}$, which imply planetary masses and minimum masses of $M_\mathrm{b}$\,=\,0.699\,$\pm$\,0.083\,M$_{\oplus}$, $M_\mathrm{c}\sin{i_\mathrm{c}}$\,=\,4.08\,$\pm$\,0.30\,M$_{\oplus}$, and $M_\mathrm{d}\sin{i_\mathrm{d}}$\,=\,5.93\,$\pm$\,0.45\,M$_{\oplus}$ for GJ\,367\,b, c, and d, respectively, whereas the RV semi-amplitudes induced by stellar activity signals at 51.3 and 138 days are of $K_\mathrm{\star,Rot}$\,=\,2.52\,$\pm$\,0.13\,m\,s$^{-1}$ and $K_\mathrm{\star,Evol}$\,=\,1.25\,$\pm$\,0.69\,m\,s$^{-1}$. The RV time series along with the best-fitting model are shown in Fig.~\ref{fig:GJ367_rv_fit}. The phase-folded RV curves for each signal are displayed in Fig.~\ref{fig:GJ367_rv_fit_folded}.

\begin{deluxetable*}{lcc}
\tablenum{3}
\tablecaption{System parameters as derived modeling the stellar signals with two sine functions.
\label{tab:fit_param}}
\tablewidth{0pt}
\tablehead{{Parameter} & \colhead{Prior} & \colhead{Derived value}}
\startdata
\noalign{\smallskip}
\textbf{GJ\,367\,b} &  &  \\ 
\textit{Model parameters} & & \\
Orbital period $P_\mathrm{orb, b}$ [days] & $\mathcal{N}$[0.3219225, 0.0000002]  & 0.3219225 $\pm$ 0.0000002  \\ 
Transit epoch $T_\mathrm{0,\mathrm{b}}$ [BJD$_\mathrm{TDB}$-2,450,000] &  $\mathcal{N}$[8544.1364,0.0004]  & 8544.13632 $\pm$ 0.00040 \\ 
$\sqrt{e_\mathrm{b}}\sin{\omega_\mathrm{\star,b}}$ & $\mathcal{U}$[-1.0,1.0]  & -0.23$^{+0.30}_{-0.23}$  \\
$\sqrt{e_\mathrm{b}}\cos{\omega_\mathrm{\star,b}}$ & $\mathcal{U}$[-1.0,1.0]  & -0.07 $\pm$ 0.13 \\
Radial velocity semi-amplitude variation $K_\mathrm{b}$ [m\,s$^{-1}$] & $\mathcal{U}$[0.00, 0.05]   & 1.10 $\pm$ 0.14 \\
\noalign{\smallskip}
\hline
\textit{Derived parameters} & &  \\
Planet mass $M_\mathrm{b}$ [M$_\oplus$]$^{(*)}$ & --  & 0.699 $\pm$ 0.083 \\
Orbit eccentricity $e_\mathrm{b}$ & --  & 0.10$^{+0.14}_{-0.07}$ \\
Argument of periastron of stellar orbit $\omega_{\star,\mathrm{b}}$\,[deg] & --  & 251$^{+23}_{-102}$  \\
\noalign{\smallskip}
\hline
\noalign{\smallskip}
\textbf{GJ\,367\,c} &  &  \\ 
\textit{Model parameters} & & \\
Orbital period $P_\mathrm{orb, c}$ [days] & $\mathcal{U}$[11.4858,11.5858]  & 11.543 $\pm$ 0.005 \\ 
\textbf{Time of inferior conjunction} T$_{0,\mathrm{c}}$ [BJD$_\mathrm{TDB}$-2,450,000] &  $\mathcal{U}$[9152.6591,9154.6591]  & 9153.46 $\pm$ 0.21 \\ 
$\sqrt{e_\mathrm{c}}\sin{\omega_\mathrm{\star,c}}$ & $\mathcal{U}$[-1,1]  & 0.38 $^{+0.10}_{-0.13}$  \\
$\sqrt{e_\mathrm{c}}\cos{\omega_\mathrm{\star,c}}$ & $\mathcal{U}$[-1,1]  & 0.27 $^{+0.11}_{-0.14}$ \\
Radial velocity semi-amplitude variation $K_\mathrm{c}$ [m\,s$^{-1}$] & $\mathcal{U}$[0.00, 0.05]   & 2.01 $\pm$ 0.15 \\
\noalign{\smallskip}
\hline
\textit{Derived parameters} & &  \\
Planet minimum mass $M_\mathrm{c}$\,$\sin{i_\mathrm{c}}$ [M$_\oplus]$ & --  & 4.08 $\pm$ 0.30 \\
Orbit eccentricity $e_\mathrm{c}$ & --  & 0.23 $\pm$ 0.07  \\
Argument of periastron of stellar orbit $\omega_{\star,\mathrm{c}}$\,[deg] & --  & 55 $\pm$ 18 \\
\noalign{\smallskip}
\hline
\noalign{\smallskip}
\textbf{GJ\,367\,d} &  &  \\ 
\textit{Model parameters} & & \\
Orbital period $P_\mathrm{orb, d}$ [days] & $\mathcal{U}$[34.0016,34.6016]  & 34.39 $\pm$ 0.06 \\ 
\textbf{Time of inferior conjunction} $T_{0,\mathrm{d}}$ [BJD$_\mathrm{TDB}$-2,450,000] &  $\mathcal{U}$[9179.2710,9183.2710]  & 9180.90 $^{+0.70}_{-0.81}$ \\
$\sqrt{e_\mathrm{d}}\cos{\omega_\mathrm{\star,d}}$ & $\mathcal{U}$[-1,1]  & $-$0.10$^{+0.20}_{-0.18}$  \\
$\sqrt{e_\mathrm{d}}\cos{\omega_\mathrm{\star,d}}$ & $\mathcal{U}$[-1,1]  & 0.16$^{+0.16}_{-0.20}$ \\
Radial velocity semi-amplitude variation $K_\mathrm{d}$ [m\,s$^{-1}$] & $\mathcal{U}$[0.00, 0.05]   & 1.98 $\pm$ 0.15 \\
\noalign{\smallskip}
\hline
\textit{Derived parameters} & &  \\
Planet minimum mass $M_\mathrm{d}$\,$\sin{i_\mathrm{d}}$ [M$_\oplus]$ & --  & 5.93 $\pm$ 0.45 \\
Orbit eccentricity $e_\mathrm{d}$ & --  & 0.08$^{+0.07}_{-0.05}$  \\
Argument of periastron of stellar orbit $\omega_{\star,\mathrm{d}}$\,[deg] & --  & 277$^{+58}_{-242}$ \\
\noalign{\smallskip}
\hline
\noalign{\smallskip}
\textbf{Stellar activity induced RV signal} & & \\
Rotation period $P_{\star,\mathrm{Rot}}$ [days] & $\mathcal{U}$[50.0903,52.0903]  & 51.30 $\pm$ 0.13 \\
Rotation RV semi-amplitude $K_\mathrm{\star,Rot}$ [m\,s$^{-1}$] & $\mathcal{U}$[0.00, 0.05]   & 2.52 $\pm$ 0.13 \\
Active region evolution period $P_{\star,\mathrm{Evol}}$ [days] & $\mathcal{U}$[103.1797,163.1797]  & 138 $\pm$ 2 \\ 
Active region evolution RV semi-amplitude $K_\mathrm{\star,Evol}$ [m\,s$^{-1}$] & $\mathcal{U}$[0.00, 0.05]   & 1.25 $\pm$ 0.14 \\
\noalign{\smallskip}
\hline
\noalign{\smallskip}
\textbf{Additional model parameters} & &  \\
Systemic velocity $\gamma_{HARPS}$ [m\,s$^{-1}$] & $\mathcal{U}$[47.806,48.025]  & 47.91674 $\pm$ 0.00013  \\
Radial velocity jitter term $\sigma_{RV,HARPS}$ [m\,s$^{-1}$] & $\mathcal{J}$[0,100]  & 1.59 $\pm$ 0.07  \\
\noalign{\smallskip}
\enddata
\tablecomments{$\mathcal{U}[a,b]$ refers to uniform priors between $a$ and $b$; $\mathcal{N}[a,b]$ refers to Gaussian priors with mean $a$ and standard deviation $b$; $\mathcal{J}[a,b]$ refers to Jeffrey's priors between $a$ and $b$. Inferred parameters and uncertainties are defined as the median and the 68.3\% credible interval of their posterior distributions.\\
$^{(*)}$ Assuming an orbital inclination of $i_\mathrm{b}$\,=\,79.89$^{+0.87}_{-0.85}$°, from the modeling of \tess transit light curves (Sect.\,\ref{FCO}).}
\end{deluxetable*}

\clearpage
\subsection{Multi-dimensional Gaussian process approach}
\label{GP}

We also followed a multidimensional Gaussian process (GP) approach to account for the stellar signals in our RV time series \citep{2015Rajpaul}. This approach models the RVs along with time series of activity indicators assuming that the same GP, a function $G(t)$, can describe them both. The function $G(t)$ represents the projected area of the visible stellar disk that is covered by active regions at a given time. For our best GP analysis, we selected the activity indicator that shows the strongest signal in the periodograms, i.e., the DLW, and modeled the RVs alongside this activity index. We created a two-dimensional GP model via
\begin{eqnarray}
   & \mathrm{RV} = A_{\mathrm{RV}} G(t) + B_{\mathrm{RV}} \dot{G}(t), \\
   & \mathrm{DLW} = A_{\mathrm{DLW}} G(t)
\label{eq:1}
\end{eqnarray}
The amplitudes $A_{\mathrm{RV}}$, $B_{\mathrm{RV}}$, and $A_{\mathrm{DLW}}$ are free parameters, which relate the individual time series to $G(t)$. The RV data are modeled as a function of $G(t)$ and its time derivative $\dot{G}(t)$, since they depend both on the fraction of the stellar surface covered by active regions, and on how these regions evolve and migrate on the disk. The DLW, which measures the width of the spectral lines, has been proven to be a good tracer of the fraction of the surface covered by active regions, and is thus expected to be solely proportional to $G(t)$ \citep{2022Zicher}. For our covariance matrix, we used a quasi-periodic kernel
\small
\begin{eqnarray}
    \gamma(t_{i},t_{j}) = \exp\left[- \frac{\sin{^2}\left[\pi(t_{i}-t_{j})/P_\mathrm{GP}\right]}{2\lambda_\mathrm{p}^{2}} - \frac{(t_{i}-t_{j})^{2}}{2\lambda_\mathrm{e}^{2}}\right]
    \label{eq_3}
\end{eqnarray}
\normalsize
and its derivatives \citep{2022Barragan, 2015Rajpaul}.  The parameter P$_\mathrm{GP}$ in Eq.\,\ref{eq_3} is the characteristic period of the GP, which is interpreted as the stellar rotation period; $\lambda_\mathrm{p}$ is the inverse of the harmonic complexity, which is associated with the distribution of active regions on the stellar surface \citep{2015Aigrain}; $\lambda_\mathrm{e}$ is the long-term evolution timescale, i.e., the active region lifetime on the stellar surface. We performed the multidimensional GP regression using \pyaneti \citep[described in][]{2022Barragan}, adding three Keplerians to account for the Doppler reflex motion of the three planets, as described in Sect.~\ref{RV_analyis}.

We performed an MCMC analysis setting informative Gaussian priors based on the transit ephemeris of the innermost planet  GJ\,367\,b, as derived in Sect.~\ref{FCO}, and uniform priors for all the remaining parameters. We also used uniform priors to sample for the multidimensional GP hyper-parameters. We included a jitter term to the diagonal of the covariance for each time series. 

We performed our fit with 500 Markov chains to sample the parameter space. 
The posterior distributions were created using the last 5000 iterations of the converged chains with a thin factor of 10, leading to a distribution of 250,000 data points for each fitted parameter. The chains were initialized at random values within the priors ranges. This ensured a homogeneous sampling of the parameter space.

Priors and results are listed are listed in Table~\ref{tab:GP_fit_param}. Planets GJ\,367\,b, c, and d are significantly detected in the \harps RV time series with Doppler semi-amplitudes of $K_\mathrm{b}$\,=\,0.86\,$\pm$\,0.15\,m\,s$^{-1}$, $K_\mathrm{c}$\,=\,1.99\,$\pm$\,0.17\,m\,s$^{-1}$, and $K_\mathrm{d}$\,=\,2.03\,$\pm$\,0.16\,m\,s$^{-1}$, respectively. These imply planetary masses and minimum masses of $M_\mathrm{b}$\,=\, 0.546\,$\pm$\,0.093\,M$_{\oplus}$, $M_\mathrm{c}\sin{i_\mathrm{c}}$\,=\,4.13\,$\pm$\,0.36\,M$_{\oplus}$, and $M_\mathrm{d}\sin{i_\mathrm{d}}$\,=\, 6.03\,$\pm$\,0.49\,M$_{\oplus}$.
The resulting GP hyper-parameters are P$_\mathrm{GP}$\,=\,53.67$^{+0.65}_{-0.53}$\,days, $\lambda_\mathrm{p}$\,=\,0.44\,$\pm$\,0.05, and $\lambda_\mathrm{e}$\,=\,114\,$\pm$\,19 \,days. The characteristic period P$_\mathrm{GP}$ is in agreement with the stellar rotation period, as discussed in Sect.\,\ref{sec:star_rot}, as well as the long-term evolution timescale $\lambda_\mathrm{e}$, which is in agreement with the long-period signal found in the analyses of Sects.\,\ref{sec:Frequency_analysis} and \ref{RV_analyis}.  

\begin{figure*}[!t]
\centering
\includegraphics[width=1.05\textwidth]{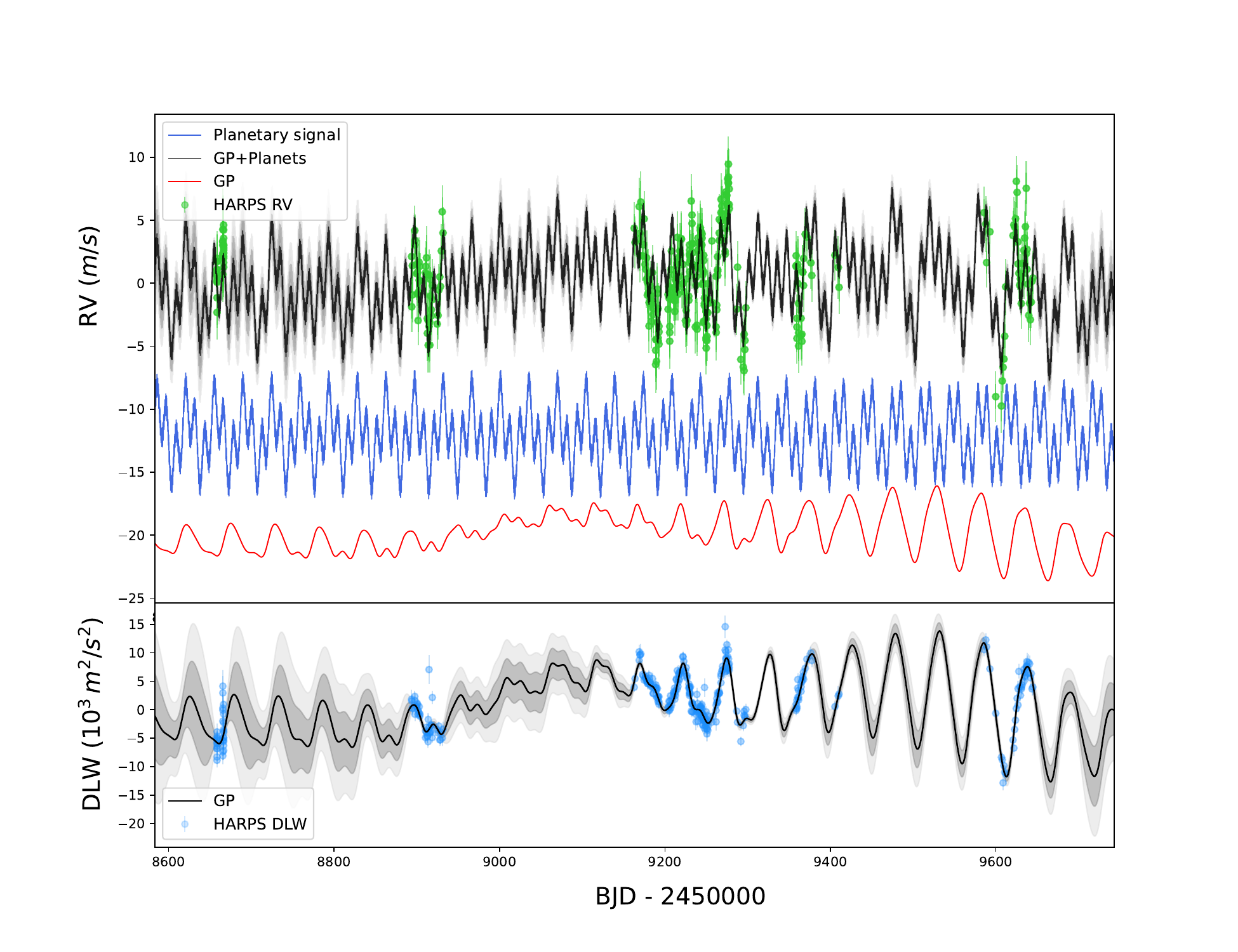}
\caption{RV and differential line width (DLW) time series with best-fitting models from the multi-GP (solid black lines) and the 1$\sigma$ and 2$\sigma$ credible intervals of the corresponding GP models (light gray shaded areas). The upper panel shows the RV data with the full model in black, whilst the planetary signal (blue), and stellar (red) inferred models are shown with a vertical offset for clarity. The lower panel shows the DLW along with the stellar inferred model. Data are shown with filled circles, with their nominal error bars and semitransparent error bar extensions accounting for the inferred jitter term. 
     \label{fig:GP_GJ367}
        }
\end{figure*}

\begin{figure*}
\gridline{\fig{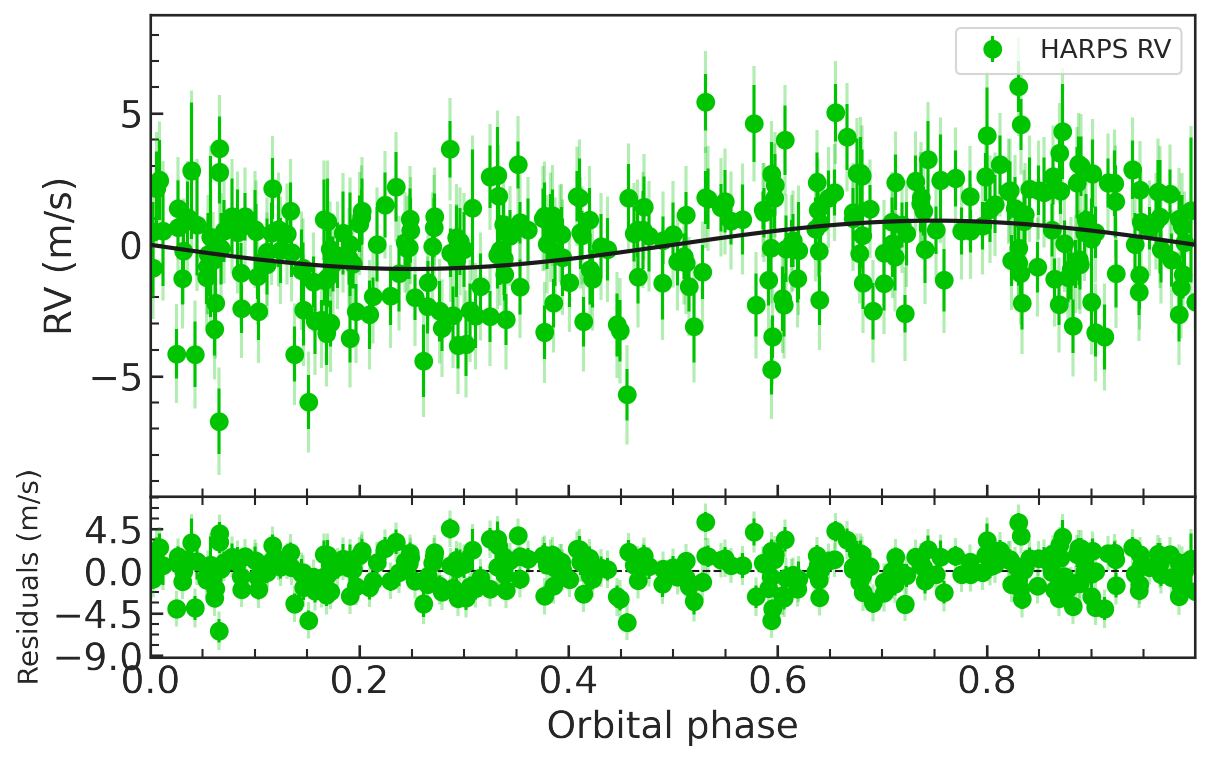}{0.45\textwidth}{(a)}
          \fig{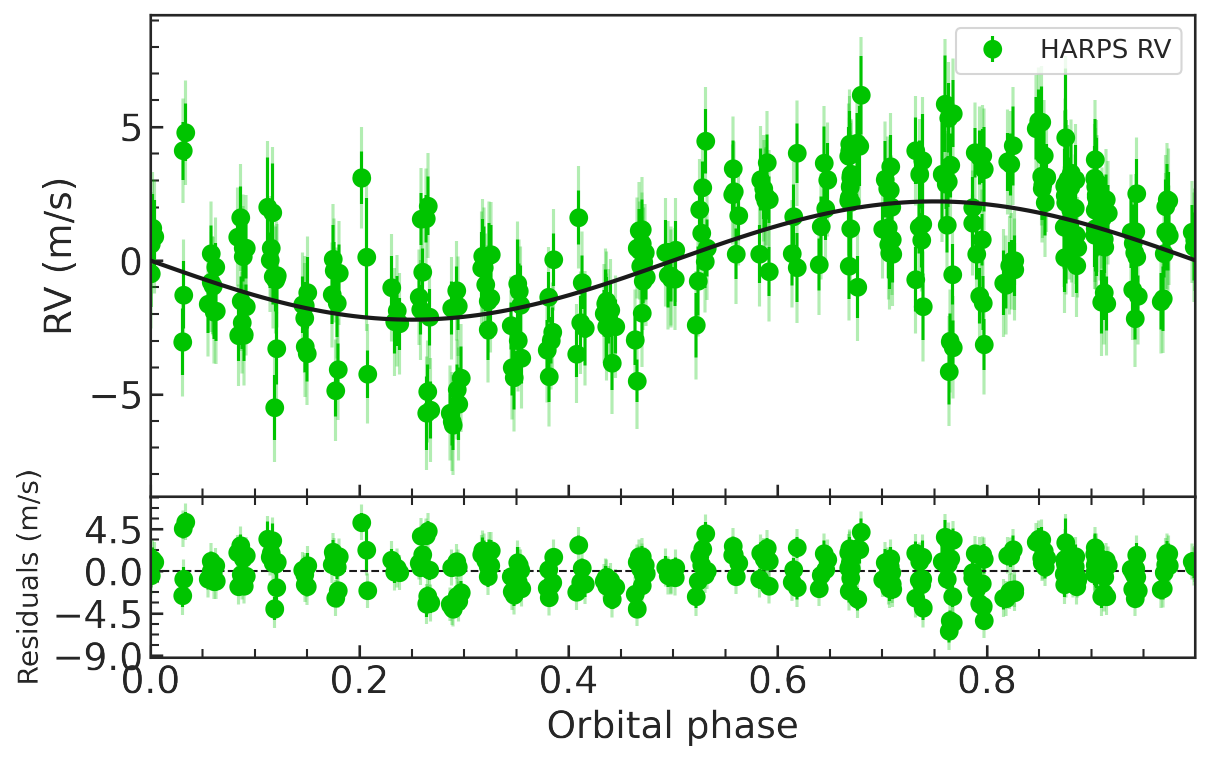}{0.45\textwidth}{(b)}}
\gridline{\fig{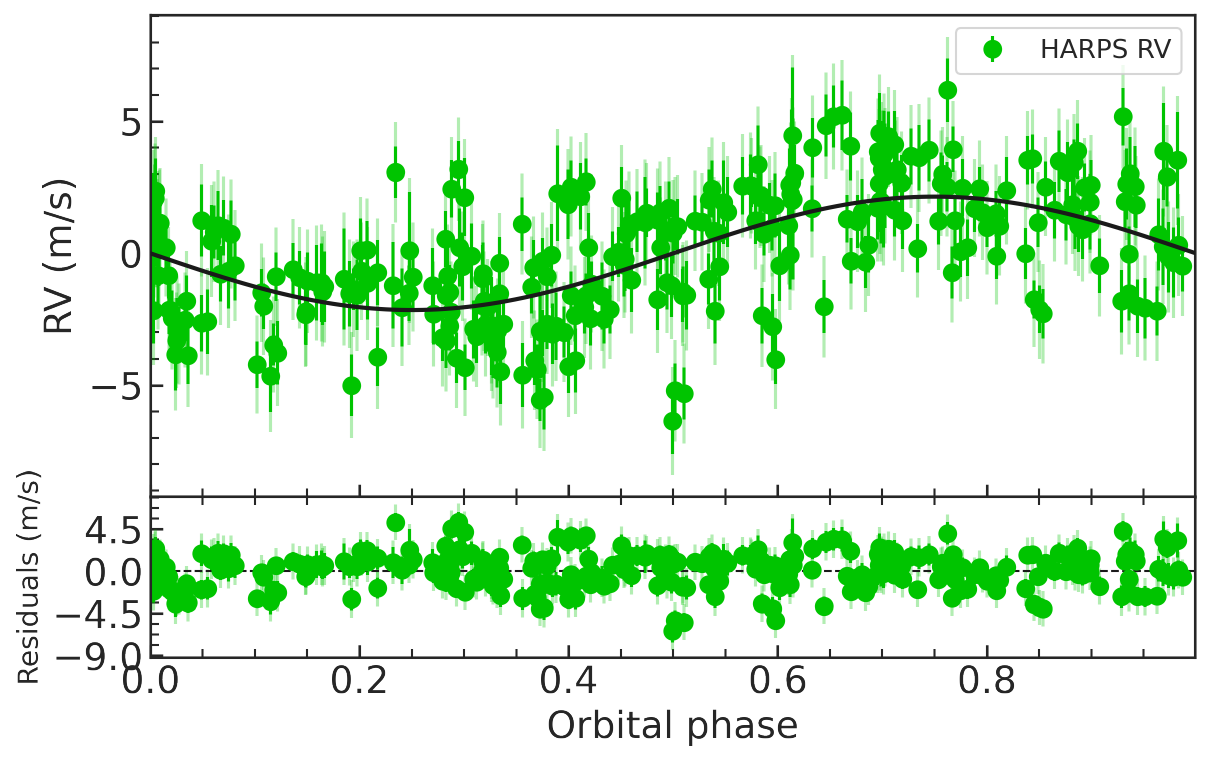}{0.45\textwidth}{(d)}}
\caption{Phase-folded RVs curve of GJ\,367\,b (a), GJ\,367\,c (b), and GJ\,367\,d (c). Data are shown as filled green circles with the error bars and the light-green error bars accounting for the inferred RV jitter term. 
     \label{fig:GP_GJ367_folded}
        }
\end{figure*}

Figure~\ref{fig:GP_GJ367} shows the RV and DLW time series, along with the inferred models, whereas Fig.~\ref{fig:GP_GJ367_folded} displays the phase-folded RV curves of the three planets and the best-fitting models.

\begin{deluxetable*}{lcc}
\tablenum{4}
\tablecaption{System parameters as derived modeling the stellar signals with a GP.
\label{tab:GP_fit_param}}
\tablewidth{0pt}
\tablehead{{Parameter} & \colhead{Prior} & \colhead{Derived value}}
\startdata
\textbf{GJ\,367\,b} &  &  \\  
\textit{Model parameters} & & \\
Orbital period $P_{\mathrm{orb,b}}$ [days] & $\mathcal{N}$[0.3219225, 0.0000002]  & 0.3219225\,$\pm$\,0.0000002  \\
Transit epoch T$_{0,\mathrm{b}}$ [BJD$_\mathrm{TDB}-$2,450,000] &  $\mathcal{N}$[8544.1364, 0.0004]  & 8544.13631\,$\pm$\,0.00038 \\ 
$\sqrt{e_{b}}\sin{\omega_{\star,\mathrm{b}}}$ & $\mathcal{U}$[$-$1.0,1.0]  & 0.13$^{+0.29}_{-0.32}$ \\
$\sqrt{e_{b}}\cos{\omega_{\star,\mathrm{b}}}$ & $\mathcal{U}$[$-$1.0,1.0] & $-$0.05\,$\pm$\,0.21 \\
Radial velocity semi-amplitude variation $K_\mathrm{b}$ [m\,s$^{-1}$] & $\mathcal{U}$[0.00, 0.05]   & 0.86\,$\pm$\,0.15 \\
\hline
\textit{Derived parameters} & &  \\
Planet mass $M_\mathrm{b}\,[M_\oplus]$$^{(*)}$ & --  & 0.546\,$\pm$\,0.093 \\
Orbit eccentricity $e_\mathrm{b}$ & --  & 0.10$^{+0.13}_{-0.07}$ \\
Argument of periastron of stellar orbit $\omega_{\star,\mathrm{b}}$\,[deg] & --  & 71$^{+60}_{-173}$ \\
\hline         
\textbf{GJ\,367\,c} &  &  \\
\textit{Model parameters} & & \\
Orbital period $P_{\mathrm{orb,c}}$ [days] & $\mathcal{U}$[11.4, 11.6]  & 11.5301\,$\pm$\,0.0078 \\ 
\textbf{Time of inferior conjunction} $T_{0,\mathrm{c}}$ [BJD$_\mathrm{TDB}-$2,450,000] &  $\mathcal{U}$[9153.0, 9155.0]  & 9153.84\,$\pm$\,0.30 \\ 
$\sqrt{e_\mathrm{c}}\sin{\omega_{\star,\mathrm{c}}}$ & $\mathcal{U}$[$-$1.0,1.0]  & $-$0.11$^{+0.23}_{-0.20}$ \\
$\sqrt{e_\mathrm{c}}\cos{\omega_{\star,\mathrm{c}}}$ & $\mathcal{U}$[$-$1.0,1.0]  & 0.14$^{+0.19}_{-0.15}$ \\
Radial velocity semi-amplitude variation $K_\mathrm{c}$ [m\,s$^{-1}$] & $\mathcal{U}$[0.00, 0.05]   & 1.99\,$\pm$\,0.17 \\
\hline
\textit{Derived parameters} & &  \\
Planet minimum mass $M_\mathrm{c}\,\sin{i_\mathrm{c}}[M_\oplus]$ & --  & 4.13\,$\pm$\,0.36 \\
Orbit eccentricity $e_\mathrm{c}$ & --  & 0.09\,$\pm$\,0.07 \\
Argument of periastron of stellar orbit $\omega_{\star,\mathrm{c}}$\,[deg] & --  & $-$34$^{+74}_{-54}$ \\
\hline
\textbf{GJ\,367\,d} &  &  \\
\textit{Model parameters} & & \\
Orbital period $P_{\mathrm{orb,d}}$ [days] & $\mathcal{U}$[30.0,40.0]  & 34.369\,$\pm$\,0.073 \\ 
\textbf{Time of inferior conjunction} $T_{0,\mathrm{d}}$ [BJD$_\mathrm{TDB}-$2,450,000] &  $\mathcal{U}$[9173.0, 9180.0]  & 9181.82\,$\pm$\,1.10 \\ 
$\sqrt{e_\mathrm{d}}\sin{\omega_{\star,\mathrm{d}}}$ & $\mathcal{U}$[$-$1.0,1.0]  & $-$0.09\,$\pm$\,0.19 \\
$\sqrt{e_\mathrm{d}}\cos{\omega_{\star,\mathrm{d}}}$ & $\mathcal{U}$[$-$1.0,1.0]  & $-$0.30$^{+0.20}_{-0.13}$ \\
Radial velocity semi-amplitude variation $K_\mathrm{d}$ [m\,s$^{-1}$] & $\mathcal{U}$[0.00, 0.05]   & 2.03\,$\pm$\,0.16 \\
\hline
\textit{Derived parameters} & &  \\
Planet minimum mass $M_\mathrm{d}\,\sin{i_\mathrm{d}}[M_\oplus]$ & --  & 6.03\,$\pm$\,0.49 \\
Orbit eccentricity $e_\mathrm{d}$ & --  & 0.14\,$\pm$\,0.09 \\
Argument of periastron of stellar orbit $\omega_{\star,\mathrm{d}}$\,[deg] & --  & $-$126$^{+287}_{-38}$ \\
\hline
\textbf{Additional model parameters} & &  \\
Characteristic period of the GP  $P_\mathrm{GP}$ [days] & $\mathcal{U}$[35.0,65.0]  & 53.67$^{+0.65}_{-0.53}$  \\
Inverse of the harmonic complexity $\lambda_\mathrm{p}$ & $\mathcal{U}$[0.1,3.0]  & 0.44\,$\pm$\,0.05 \\
Long-term evolution timescale $\lambda_\mathrm{e}$ [days] & $\mathcal{U}$[1,300]  & 114\,$\pm$\,19 \\
Amplitude $A_{\mathrm{RV}}$ [km\,s$^{-1}$] & $\mathcal{U}$[0.0,0.005]  & 0.0016\,$\pm$\,0.0004  \\
Amplitude $B_{\mathrm{RV}}$ [km\,s$^{-1}$] & $\mathcal{U}$[$-$0.05,0.05]  & 0.009$^{+0.0025}_{-0.0019}$ \\
Amplitude $A_{\mathrm{DLW}}$ & $\mathcal{U}$[0.0,30.0]  &  6.31$^{+1.42}_{-1.02}$   \\
Systemic velocity $\gamma_{\mathrm{HARPS}}$ [km\,s$^{-1}$] & $\mathcal{U}$[47.40,48.42]  & 47.9168\,$\pm$\,0.0005 \\
Offset DLW [10$^3$\,m$^{2}$\,s$^{-2}$ ]& $\mathcal{U}$[$-$16.90,14.11]  & $-$0.11$^{+2.02}_{-1.98}$ \\
Radial velocity jitter term $\sigma_{\mathrm{HARPS}}$ [m\,s$^{-1}$] & $\mathcal{J}$[0,100]  & 1.83\,$\pm$\,0.08  \\
DLW jitter term $\sigma_{\mathrm{DLW}}$ [m$^{2}$\,s$^{-2}$] & $\mathcal{J}$[0,1000]  & 378$^{+145}_{-335}$
\enddata
\tablecomments{$\mathcal{U}[a,b]$ refers to uniform priors between $a$ and $b$; $\mathcal{N}[a,b]$ refers to Gaussian priors with mean $a$ and standard deviation $b$; $\mathcal{J}[a,b]$ refers to Jeffrey's priors between $a$ and $b$. Inferred parameters and uncertainties are defined as the median and the 68.3\% credible interval of their posterior distributions.\\
$^{(*)}$ Assuming an orbital inclination of $i_\mathrm{b}$\,=\,79.89$^{+0.87}_{-0.85}$°, from the modeling of \tess transit light curves (Sect.\,\ref{FCO}).}
\end{deluxetable*}

\section{Discussion}
\label{sec:Discussion}

The three techniques used to determine the mass of GJ\,367\,b give results that are consistent to within $\sim$1$\sigma$. We adopted the results from the FCO method, which gives a planetary mass of $M_\mathrm{b}$\,=\,0.633\,$\pm$\,0.050~M$_{\oplus}$ (7.9\% precision). Our result differs by about $\sim$1$\sigma$ from the mass of $M_\mathrm{b}$\,=\,0.546\,$\pm$ 0.078~M$_{\oplus}$ reported by \cite{2021Lam}, which was also derived using the FCO method applied on 20 \harps\ data chunks that do not entirely cover the orbital phase of the USP planet, as opposed to our 96 chunks. 
We found that GJ\,367\,b has a radius of $R_\mathrm{b}$\,=\,0.699\,$\pm$\,0.024\,R$_{\oplus}$ (3.4\% precision), consistent with the value of $R_\mathrm{b}$\,=\,0.718\,$\pm$\,0.054\,R$_{\oplus}$ from \cite{2021Lam}, but more precise, thanks to the additional \tess\ photometry and increased cadence. \cite{2021Lam} reported a density of $\rho_\mathrm{b}$\,=\,8.1\,$\pm$\,2.2~g\,cm$^{-3}$. The higher mass but similar radius measured in this work make the USP planet denser, with a ultra-high density of $\rho_\mathrm{b}$\,=\,10.2\,$\pm$\,1.3~g\,cm$^{-3}$.

\begin{figure*}
\centering
\includegraphics[width=0.75\linewidth]{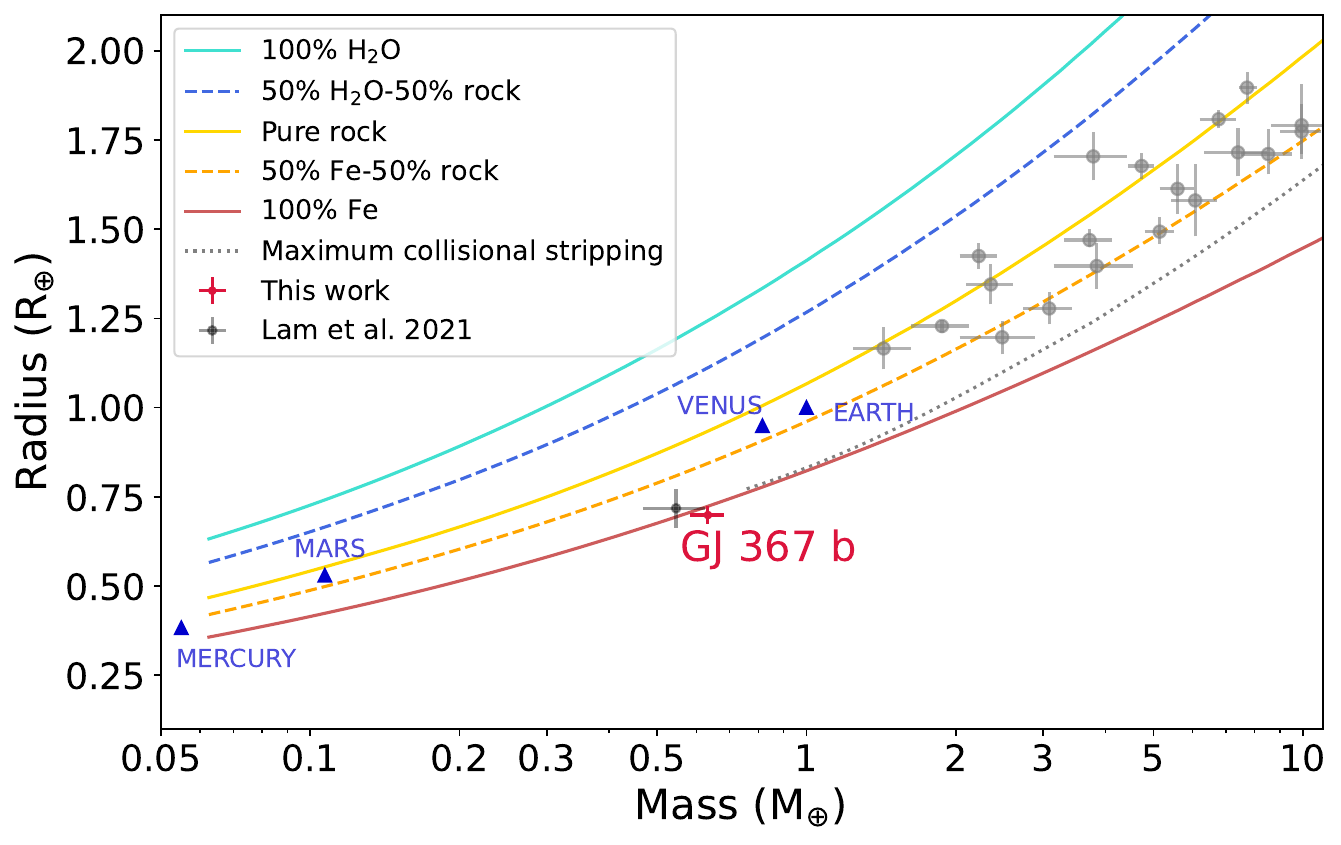}
    \caption{Mass-radius diagram of small USP planets (P\,$<$\,1\,days, R\,$<$\,2\,R$_{\oplus}$, M\,$<$\,10\,M$_{\oplus}$) with masses and radii known with a precision better than 20\%, as retrieved from the Transiting Extrasolar Planet Catalogue \citep{2011Southworth}. From bottom to top, the solid and dashed curves are theoretical models from \cite{2016Zeng}. 
    We highlighted GJ~367 b with a red dot and the previous position on the diagram with a black dot \citep{2021Lam}. }
    \label{MR_diagram}
\end{figure*}

GJ\,367\,b belongs to the handful of small USP planets ($R_\mathrm{p}$\,$<$\,2~R$_{\oplus}$, $M_\mathrm{p}$\,$<$\,10~M$_{\oplus}$) whose masses and radii are known with a precision better than 20\%. Figure~\ref{MR_diagram} shows the mass-radius diagram for small USP planets along with the theoretical composition models for rocky worlds \citep{2016Zeng}. GJ\,367\,b is the smallest and densest USP planet known to date. The position of the planet on the mass-radius diagram suggests that its composition is dominated by iron. 
Taking into account its mean density, GJ\,367\,b leads the class of super-Mercuries, namely, extremely dense planets containing an excess of iron, analogous to Mercury: K2-229\,b \citep{Santerne_2018}, K2-38\,b \citep{2020Toledo}, K2-106\,b \citep{Guenther_2017}, Kepler-107\,c \citep{Bonomo_2019}, Kepler-406\,b \citep{2014Marcy}, HD 137496\,b \citep{Azevedo_Silva_2022}, HD 23472\,b \citep{Barros_2022}, and TOI-1075\,b \citep{2022Essack}. 

The two non-transiting planets GJ\,367\,c and GJ\,367\,d have orbital periods of $\sim$11.5 d and 34.4 days, respectively, and minimum masses of $M_\mathrm{c}$\,$\sin{i_\mathrm{c}}$\,=\,4.13\,$\pm$\,0.36~M$_{\oplus}$ and $M_\mathrm{d}$\,$\sin{i_\mathrm{d}}$\,=\,6.03\,$\pm$\,0.49~M$_{\oplus}$, as derived adopting the multidimensional GP approach to model stellar activity (Sect.\,\ref{GP}). We note that our minimum mass determinations are in very good agreement with those of $M_\mathrm{c}$\,$\sin{i_\mathrm{c}}$\,=\,4.08\,$\pm$\,0.30~M$_{\oplus}$ and $M_\mathrm{d}$\,$\sin{i_\mathrm{d}}$ = 5.93\,$\pm$\,0.45~M$_{\oplus}$ we determined modeling stellar activity with two sinusoidal components (Sect.~\ref{RV_analyis}). 

If the orbits of GJ\,367\,b, c, and d were coplanar ($i_\mathrm{b}$\,=\,$i_\mathrm{c}$\,=\,$i_\mathrm{d}$\,=\,79.9°), planets c and d would have true masses of $M_\mathrm{c}$\,=\,4.19\,$\pm$\,0.35~M$_{\oplus}$ and $M_\mathrm{d}$\,=\,6.12\,$\pm$\,0.48~M$_{\oplus}$, respectively.
Using the mass-radius relation for small rocky planets from \citet{Otegi_2020}, we found that GJ\,367\,c and d are expected to have radii of $\sim$1.6 and $\sim$1.7~R$_{\oplus}$, respectively, making them two super-Earths with mean densities of $\sim$6~g\,cm$^{-3}$. 
We searched the \tess\ light curves for possible transits of the two outer companions with the DST code (Sect.~\ref{tess_obs}) masking out the transits of the UPS planet, but we found no other significant transit signals. Under the assumption that the orbits of the three planets are coplanar, the impact parameters of planets c and d would be $b_\mathrm{c}$\,$\approx$\,6 and $b_\mathrm{d}$\,$\approx$\,13, respectively. This would account for the non-detection of the transit signals of GJ\,367\,c and GJ\,367\,d in the \tess\ light curves. In this scenario, GJ\,367\,c and d would transit their host star only if their radii were unphysically large, i.e. $R_c\,>1.9\, R_{\odot}$ and $R_d\,>5\,R_{\odot}$.

The other configuration for the outer planets to transit is if they are mutually inclined with planet b ($i_\mathrm{c}\,\neq\,i_\mathrm{b};i_\mathrm{d}\,\neq\,i_\mathrm{b}$). This is not an unusual architecture for systems with USP planets \citep{2018Dai}.
With minimum masses of 4.13 $M_{\oplus}$ and 6.03 $M_{\oplus}$, GJ\,367\,c and d  are expected to have a minimum radius of $\sim$1.5\,$R_{\oplus}$ given the maximum collisional stripping limit \citep{marcus_collisional_2009,marcus_minimum_2010}. The inclination of their orbits should be larger than $\sim$86.5° to produce non-grazing transits. If the two planets had radii of 1.5\,$R_{\oplus}$, the transit depths would be $\sim$900\,ppm. For 2.5\,R$_{\oplus}$, the non grazing transit depth is expected to be $\sim$2500\,ppm. Similarly, transit depth of a 4\,$R_{\oplus}$ planet is $\sim 6400$ ppm. The  rms of the \tess\ light curve of GJ\,367 is approximately 500\,ppm. The expected transit times of planets c and d also fall well within the baseline of the \tess\ data. Therefore, GJ\,367\,c and d would be easily detected if they produced non-grazing transits.

The presence of two additional planetary companions to GJ\,367\,b is in line with the tendency of USP planets to belong to multi-planet systems \citep{2021Dai}. While the origin of USP planets is debated, it is likely that the presence of outer companions could be responsible for the migration processes that carried USP planets to their current positions \citep{2019Pu, 2019Petrovich}, which would lie inside the magnetospheric cavity of a typical protoplanetary disk. In these models, planet migration occurs after the dissipation of the protoplanetary disk, by a combination of eccentricity forcing from the planetary companions and tidal dissipation in the innermost planet's interior, although the precise dynamical forcing mechanisms are debated. \cite{Serrano2022} showed, for TOI-500, that the low-eccentricity secular forcing proposed by \cite{2019Pu} can explain the migration of that system's USP planet from a formation radius of $\sim$0.02\,au to its observed location. For GJ\,367\,b, we tested this migration scenario with similar initial conditions (USP planet starting at $0.02$\,au, initial eccentricities of all planets set to $0.2$), and found only modest migration of planet~b, from $0.02$\,au to $\sim$0.01\,au, short of the observed $0.007$\,au. This would support an alternative migration history, such as by high-eccentricity secular chaos \citep{2019Petrovich}. However, we note that GJ\,367\,c and GJ\,367\,d lie close to a 3:1 mean motion commensurability, having a period ratio of 2.98, and that the dynamics may be affected by the 3:1 resonance, which can provide additional eccentricity forcing in the system. A deeper analysis would be needed to draw definite conclusions on the formation, migration, and dynamics of the system. The GJ\,367 system is thus an excellent target for studying planetary system formation and evolution scenarios.

\subsection{Internal structure of GJ\,367\,b}
\label{Internal structure}

\begin{figure*}[t!]
\centering
\includegraphics[width =0.6\textwidth]{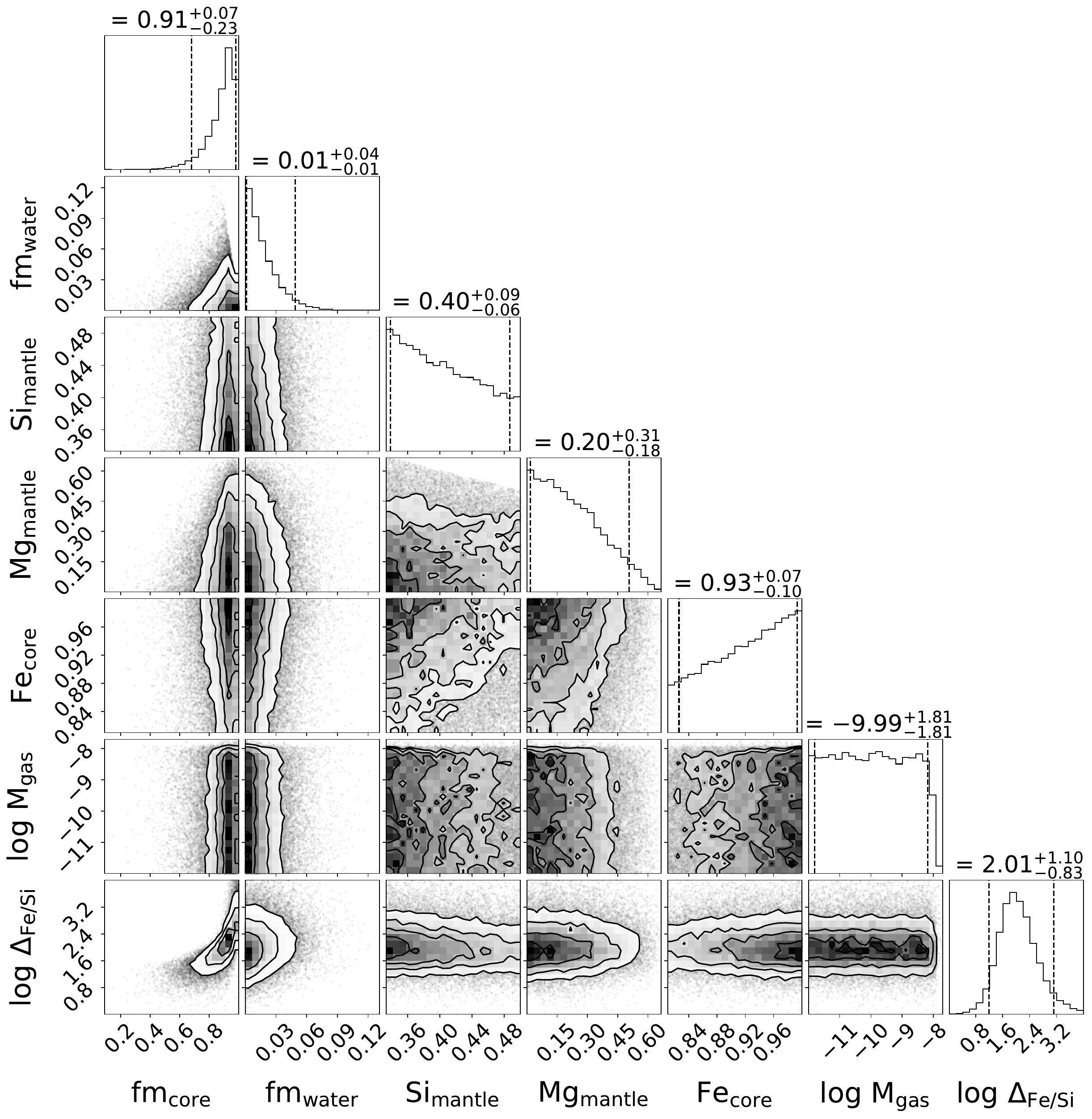}
\caption{Results of a Bayesian analysis of the internal structure of GJ\,367\,b. The depicted internal structure parameters are: the mass fractions of the inner iron core and a possible water layer with respect to the solid planet without a possible H/He layer), the molar fractions of Si and Mg within the silicate mantle layer, the molar fraction of iron in the inner core, the logarithm with base 10 of the mass of the H/He layer in Earth masses, and the shift in the Fe/Si ratio of the planet with respect to the Fe/Si ratio in the host star, again in a logarithmic scale. The titles in each column show the median of the posterior distributions and the 5th and 95th percentiles.
\label{fig:GJ367_intstruct}
}
\end{figure*}

Given the high precision of both the derived mass and radius, we used a Bayesian analysis to infer the internal structure of GJ\,367\,b. We followed the method described in \citet{2021Leleu}, which is based on \citet{2017Dorn}. Prior to presenting the results of our modeling, we provide the reader with a brief overview over the used model.

Modeling the interior of an exoplanet is a degenerate problem: there is a multitude of different compositions that could explain the observed planetary density. This is why a Bayesian modeling approach is used, with the goal of finding posterior distributions for the internal structure parameters. We assumed a planet that is made up of four fully distinct layers: an inner iron core, a silicate mantle, a water layer, and a gas layer made up of hydrogen and helium. In our forward model, this atmospheric layer is modeled separately from the rest of the planet following \citet{2014LopezFortney} and it is assumed that it does not influence the inner layers. While the presence of a gas and water layer is not expected given the high equilibrium temperature of GJ\,367\,b, we still included them in the initial model setup, as this is the most general way to model the planet and give us all possible compositions that could lead to the observed planetary density. As input parameters, we used the planetary and stellar parameters listed in Tables \ref{tab:1} and \ref{table:2}, i.e. the transit depth, RV semi-amplitude and period of the planet and the mass, radius, effective temperature, and metallicity of the star. In addition, we chose a prior distribution of $5 \pm 5$\,Gyr for the age of the star. 

For our Bayesian analysis, we chose a prior that is uniform in log for the gas mass fraction of the planet. For the mass fractions of the inner iron core, the silicate mantle, and the water layer (all with respect to the solid part of the planet without the H/He gas layer), our chosen prior is a distribution that is uniform on the simplex on which these three mass fractions add up to one. Additionally, we added an upper limit of 0.5 for the water mass fraction in the planet, as a planet with a pure water composition is not physical \citep{2014Thiabaud, 2014Marboeuf}. Note that choosing very different priors influences the results of the internal structure analysis.

There are multiple studies that show a correlation between the composition of planets and their host star \citep[e.g.][]{2015Thiabaud, 2021Adibekyan}, but it is not clear if this correlation is 1:1 or has a different slope. As a first step, we assumed the composition of the planet to match that of its host star exactly. Since we do not have measured values for the stellar [Si/H] and [Mg/H], we left them unconstrained and sampled the stellar parameters from [Si/H]\,=\,[Mg/H]\,$= 0^{+1}_{-1}$. However, our analysis showed that the observed mass and radius values of GJ\,367\,b are not compatible with a 1:1 relationship between the Si/Mg/Fe ratios of the planet and of these sampled synthetic host stars, as it is not possible to reach such a high density under these constraints. The same is true when assuming the steeper slope between stellar and planetary Fe/(Si+Mg) ratios found by \citet{2021Adibekyan}. We then repeated our analysis allowing the iron to silicon and iron to magnesium ratios in the planet to be up to a factor 1000 higher than in the sampled stars.

The results of this second analysis are summarized in Figure\,\ref{fig:GJ367_intstruct}. Indeed, we can see that the Fe/Si ratio in the planet (and therefore also the Fe/Mg ratio) was a factor of 10 to the power of $2.01^{+1.10}_{-0.83}$ higher than in the sampled host star. We further expect GJ\,367\,b to host no H/He envelope and no significant water layer. Conversely, we expect the mass fraction of the inner iron core with $0.91^{+0.07}_{-0.23}$ (median with 5th and 95th percentiles) to be very high.

\subsection{Formation and evolution of the ultra-high density sub-Earth GJ\,367\,b}\label{coll_strip}

It is not clear how a low-mass high-density planet like GJ\,367\,b forms. Possible pathways may include the formation out of material significantly more iron rich than thought to be normally present in protoplanetry disks. Although it is not clear if disks with such a large relative iron content specifically near the inner edge (where most of the material might be obtained from) exist \citep{dullemond_inner_2010,Aguichine2020,2021Adibekyan} 

Another possibility is the formation of a larger planet with a lower bulk density. Subsequently the planet differentiates into a denser core and less dense outer layers. These outer layers are then removed. This may be accomplished through two processes. 

(i) A first process is collisional stripping. The preferred removal of outer material in giant collisions \citep[e.g.,][]{marcus_collisional_2009,reinhardt_forming_2022} might have increased the bulk density. Indeed, this is what might have lead to the high iron fraction and therefore high density of Mercury for its small size \citep{benz_collisional_1988,benz_origin_2007}, as a Mercury with a chondritic composition would have a lower density. The amount of maximum stripping is governed by the mass, impact velocity, and impact parameter of the impactor \citep{marcus_minimum_2010,leinhardt_collisions_2012}. Preferable removal of outer layers requires the right mass ratio (close to unity), right impact parameter (close to grazing), and right relative velocity. There is also evidence that, at least in some systems, densities have been altered by collisions (Kepler-107; \citealt{Bonomo_2019}) A problem to effectively remove mass might be the removal of collision debris and the avoidance of a re-accretion of debris material onto the planet, as re-accretion would leave the bulk density largely unchanged. However this might not be such a large problem as originally thought \citep{spalding_solar_2020}.  Our measurement of the bulk density of GJ\,367\,b suggests that collisional stripping has to be remarkably effective in removing non-iron material from the planet if it is the only process at work.

(ii) A second process that might have played a role in shaping GJ\,367\,b after core formation is removal of outer layers of material facilitated by the enormous stellar radiation to which this planet is subjected. At the equilibrium temperature of 1365\,$\pm$\,32 K, material might sublimate, be uplifted, and transported away from the surface.

Of course, all of the above discussed processes could have contributed to create the nearly pure ball of iron, known as GJ\,367 b.

\section{Conclusions}
\label{sec:Conclusions}

We report refined mass and radius determinations for GJ\,367\,b, the ultra-high density, USP sub-Earth planet transiting the M-dwarf star GJ\,367 recently discovered by \citet{2021Lam}. We used new \tess observations from sectors 35 and 36 combined with 371 Doppler measurements collected as part of an intensive RV follow-up campaign with the \harps spectrograph. We derived a precise planetary mass of $M_\mathrm{b}$\,=\,0.633\,$\pm$\,0.050\,M$_{\oplus}$ (7.9\,\% precision) and a radius of $R_\mathrm{b}$\,=\,0.699\,$\pm$\,0.024\,R$_{\oplus}$ (3.4\% precision), resulting in an ultra-high density of $\rho_\mathrm{b}$ = 10.2\,$\pm$\,1.3~g\,cm$^{-3}$($\sim$13\%). According to our internal structure analysis, GJ\,367\,b is predominantly composed of iron with an iron core mass fraction of 0.91$_{-0.23}^{+0.07}$, which accounts for the aforementioned planetary density. In addition, our \harps RV follow-up observations, which span a period of nearly $\sim$3 yr, allowed us to discover two additional non-transiting small companions with orbital periods of $\sim$11.5 and 34.4 days, and minimum masses of $M_\mathrm{c}$\,$\sin{i_\mathrm{c}}$\,=\,4.13\,$\pm$\,0.36~M$_{\oplus}$ and $M_\mathrm{d}$\,$\sin{i_\mathrm{d}}$\,=\,6.03\,$\pm$\,0.49~M$_{\oplus}$. 
GJ\,367 joins the small group of well-characterized multiplanetary systems hosting a USP planet, with the inner planet GJ\,367\,b being the densest and smallest USP planet known to date. This unique multiplanetary system hosting this ultra-high density, USP sub-Earth is an extraordinary target to further investigate the formation and migration scenarios of USP systems. 

\clearpage

\small
\noindent
\textit{Acknowledgments:}
      We thank the anonymous referee for a thoughtful review of our paper and very positive feedback.
      This work was supported by the KESPRINT collaboration, an international consortium devoted to the characterization and research of exoplanets discovered with space-based missions (www.kesprint.science).
      Based on observations made with the ESO-3.6\,m telescope at La Silla Observatory under programs 1102.C-0923 and 106.21TJ.001. We are extremely grateful to the ESO staff members for their unique and superb support during the observations.
      We acknowledge the use of public \tess\ data from pipelines at the \tess\ Science Office and at the \tess\ Science Processing Operations Center. \tess data presented in this paper were obtained from the Milkulski Archive for Space Telescopes (MAST) at the Space Telescope Science Institute. The specific observations analyzed can be accessed via \dataset[DOI: 10.17909/rkwv-t847]{https://doi.org/10.17909/rkwv-t847}.
      Resources supporting this work were provided by the NASA High-End Computing (HEC) Program through the NASA Advanced Supercomputing (NAS) Division at Ames Research Center for the production of the SPOC data products. E.G. acknowledges the generous support from Deutsche Forschungsgemeinschaft (DFG) of the grant HA3279/14-1. D.G. gratefully acknowledges financial support from the \textit{Cassa di Risparmio di Torino} (CRT) foundation under Grant No. 2018.2323 ``Gaseous or rocky? Unveiling the nature of small worlds''. 
      Y.A. and J.A.E. acknowledge the support of the Swiss National Fund under grant 200020\_192038.
      C.M.P. gratefully acknowledges the support of the  Swedish National Space Agency (DNR 65/19). 
      R.L. acknowledges funding from University of La Laguna through the Margarita Salas Fellowship from the Spanish Ministry of Universities ref. UNI/551/2021-May 26, and under the EU Next Generation funds.
      K.W.F.L. was supported by Deutsche Forschungsgemeinschaft grants RA714/14-1 within the DFG Schwerpunkt SPP 1992, Exploring the Diversity of Extrasolar Planets.
      S.A. acknowledges  the support from the Danish Council for Independent Research through a grant No.2032-00230B.
      O.B. acknowledges that has received funding from the ERC under the European Union's Horizon 2020 research and innovation program (grant agreement No. 865624).
      H.J.D. acknowledges support from the Spanish Research Agency of the Ministry of Science and Innovation (AEI-MICINN) under grant 'Contribution of the IAC to the PLATO Space Mission' with reference PID2019-107061GB-C66.
      A.J.M. acknowledges support from the Swedish National Space Agency (career grant 120/19C).
      O.K. acknowledges support by the Swedish Research Council (grant agreement No. 2019-03548), the Swedish National Space Agency, and the Royal Swedish Academy of Sciences
      J.K. gratefully acknowledges the support of the Swedish National Space Agency (SNSA; DNR 2020-00104) and of the Swedish Research Council (VR: Etableringsbidrag 2017-04945).

\normalsize

\appendix
\noindent
We report here the \harps RVs, including those previously reported in \cite{2021Lam}, along with the activity indicators and line profile variation diagnostics extracted with NAIRA and \serval (\ref{table-GJ367-RV}). Time stamps are given in Barycentric Julian Date in the Barycentric Dynamical Time (BJD$_{\mathrm{TDB}}$).

\restartappendixnumbering

\begin{longrotatetable}
\begin{deluxetable*}{lrrrrrrrrrrrrrrrr}
\tablecaption{Radial velocities and spectral activity indicators.\label{table-GJ367-RV}}
\tablewidth{700pt}
\tabletypesize{\scriptsize}
\tablehead{
\colhead{BJD$_\mathrm{TBD}$} & \colhead{RV} & 
\colhead{$\sigma_\mathrm{RV}$} & \colhead{H$\alpha$} & 
\colhead{$\sigma_\mathrm{H\alpha}$} & \colhead{H$\beta$} & 
\colhead{$\sigma_\mathrm{H\beta}$} & \colhead{H$\gamma$} & 
\colhead{$\sigma_\mathrm{H\gamma}$} & \colhead{log\,R$'_{HK}$} & \colhead{$\sigma_\mathrm{log\,R'_{HK}}$} & \colhead{NaD} & \colhead{$\sigma_\mathrm{NaD}$} & \colhead{DLW} & \colhead{$\sigma_\mathrm{DLW}$} & \colhead{CRX} & \colhead{$\sigma_\mathrm{CRX}$} \\ 
\colhead{-2450000} & \colhead{($\mathrm{km\,s^{-1}}$)} & \colhead{($\mathrm{km\,s^{-1}}$)} & \colhead{(Ang)} & 
\colhead{(Ang)} & \colhead{(Ang)} & \colhead{(Ang)} &
\colhead{(Ang)} & \colhead{(Ang)} & \colhead{} & \colhead{} & \colhead{(Ang)} & \colhead{(Ang)} & \colhead{($\mathrm{m^{2}\,s^{-2}}$)} & \colhead{($\mathrm{m^{2}\,s^{-2}}$)} & \colhead{($\mathrm{m\,s^{-1}}$)} & \colhead{($\mathrm{m\,s^{-1}}$)}
} 
\startdata
  8658.45473 &        47.91637 &        0.00084 &        0.06472 &        0.00012 &        0.05217 &        0.00026 &        0.11412 &        0.00081 &         -5.222 &        0.077 &        0.01038 &        0.00007 &           -7.9 &          1.3 &          -10.2 &          11.1\\
  8658.46195 &        47.91814 &        0.00082 &        0.06421 &        0.00011 &        0.05180 &        0.00025 &        0.11147 &        0.00078 &         -5.180 &        0.077 &        0.01040 &        0.00007 &          -10.0 &          1.2 &            3.4 &           9.3\\
  8658.46946 &        47.91502 &        0.00085 &        0.06424 &        0.00012 &        0.05180 &        0.00026 &        0.11413 &        0.00084 &         -5.221 &        0.077 &        0.01045 &        0.00007 &           -6.8 &          1.2 &          -17.2 &          10.7\\
  8658.47642 &        47.91611 &        0.00079 &        0.06417 &        0.00011 &        0.05257 &        0.00025 &        0.11363 &        0.00078 &         -5.130 &        0.077 &        0.01041 &        0.00007 &           -7.1 &          1.3 &            2.8 &          10.9\\
  8658.48401 &        47.91860 &        0.00090 &        0.06454 &        0.00012 &        0.05315 &        0.00028 &        0.11124 &        0.00088 &         -5.139 &        0.077 &        0.01040 &        0.00008 &           -7.5 &          1.4 &            3.5 &          11.0\\
  8658.49105 &        47.91862 &        0.00087 &        0.06499 &        0.00012 &        0.05417 &        0.00027 &        0.11592 &        0.00086 &         -5.136 &        0.077 &        0.01054 &        0.00007 &           -6.1 &          1.3 &            4.4 &          10.9\\
  8658.49842 &        47.91780 &        0.00086 &        0.06548 &        0.00012 &        0.05403 &        0.00027 &        0.11601 &        0.00088 &         -5.151 &        0.077 &        0.01074 &        0.00007 &           -7.9 &          1.1 &          -11.5 &          10.4\\
  8658.50565 &        47.91753 &        0.00091 &        0.06500 &        0.00012 &        0.05374 &        0.00029 &        0.11864 &        0.00092 &         -5.137 &        0.077 &        0.01066 &        0.00008 &           -5.0 &          1.5 &            9.4 &          11.2\\
  8658.51297 &        47.91784 &        0.00089 &        0.06466 &        0.00012 &        0.05302 &        0.00028 &        0.11529 &        0.00091 &         -5.175 &        0.077 &        0.01057 &        0.00008 &           -9.3 &          1.3 &          -15.4 &          10.6\\
  8658.52049 &        47.91557 &        0.00083 &        0.06464 &        0.00011 &        0.05272 &        0.00026 &        0.11484 &        0.00085 &         -5.151 &        0.077 &        0.01069 &        0.00007 &           -5.7 &          1.3 &            6.2 &          11.1\\
  ... &       ... &        ... &        ... &        ... &        ... &        ... &        ... &        ... &         ... &        ... &        ... &        ... &            ... &          ... &            ... &           ...\\
\enddata
\tablecomments{The entire RV data set is available in its entirety in machine-readable form. Only a portion of this table is shown here to demonstrate its form and content.}
\end{deluxetable*}
\end{longrotatetable}

\bibliography{main_GJ367}{}
\bibliographystyle{aasjournal}

\end{document}